\documentclass[aps,onecolumn,11pt,floatfix,altaffilletter,tightenlines,showpacs,showkeys,notitlepage,nofootinbib]{revtex4-1}
\pdfoutput=1
\usepackage[colorlinks=true,citecolor=blue,linkcolor=blue]{hyperref}
\usepackage[normalem]{ulem}
\usepackage{amsmath,amssymb}
\usepackage{epsfig}
\usepackage{graphicx}               
\usepackage{url}
\usepackage{color}
\usepackage{multirow}
\usepackage{placeins}
\usepackage[dvipsnames]{xcolor}
\usepackage{epstopdf}
\usepackage{extarrows}
\usepackage{relsize}
\usepackage{siunitx}
\usepackage{csquotes}
\usepackage{tikz}
\usepackage[capitalize]{cleveref}
\usepackage{subfigure} 
\usepackage{lipsum}
\usepackage{gensymb}

\usetikzlibrary{trees}
\usetikzlibrary{decorations.pathmorphing}
\usetikzlibrary{decorations.markings}

\definecolor{jblue}  {RGB}{20,50,100}
\definecolor{npurple}  {RGB} {153, 51, 204}
\definecolor{wred}   {RGB}{217,0,56}
\definecolor{white}   {RGB}{255,255,255}
\definecolor{korange}   {RGB}{235, 80,  43}
\definecolor{korange2}   {RGB}{245, 100,  63}
\definecolor{kyelloworange}   {RGB}{255, 210,  110}
\definecolor{kyelloworange2}   {RGB}{240, 170,  90}
\definecolor{kred}   {RGB}{204,  102, 153}
\definecolor{kpurple}   {RGB}{153,  61, 190}
\definecolor{kpurplelight}   {RGB}{213,  161, 230}
\definecolor{red}{rgb}{1.0, 0, 0}

\newcommand{\tr}{\text{Tr}}

\newcommand{\MSbar}[0]{\overline{\text{MS}}}

\newcommand{\dd}{\mbox{d}}
\newcommand{\E}{\mathrm{e}}
\newcommand{\I}{\mathrm{i}}
\newcommand{\hc}{\text{h$.$c$.$}}
\newcommand{\ie}{i.e.~}
\newcommand{\eg}{e.g.~}
\newcommand{\cf}{cf.~}
\newcommand{\del}{\partial}
\newcommand{\transpose}{\intercal}

\newcommand{\tinytext}[1]{\text{\tiny{#1}}}
\newcommand{\fineq}[1]{\;#1}

\newcommand{\MPl}{M_\tinytext{Pl}}
\newcommand{\nn}{\nonumber}
\newcommand{\dr}{r}
\newcommand{\symh}[2]{\hspace{#1}{#2}\hspace{#1}}

\allowdisplaybreaks

\bibpunct{[}{]}{,}{n}{}{,}


\begin{document}

\title{Gravitational Waves as a Probe of Left-Right Symmetry Breaking}

\author{Vedran Brdar}   \email{vbrdar@mpi-hd.mpg.de}
\author{Lukas Graf}   \email{lukas.graf@mpi-hd.mpg.de}
\author{Alexander J. Helmboldt} \email{alexander.helmboldt@mpi-hd.mpg.de}
\author{Xun-Jie Xu} \email{xunjie.xu@mpi-hd.mpg.de}
\affiliation{Max-Planck-Institut f\"ur Kernphysik,
       69117~Heidelberg, Germany}

\begin{abstract}
\noindent
Left-right symmetry at high energy scales is a well-motivated extension of the Standard Model. In this paper we consider a typical minimal scenario in which it gets spontaneously broken by scalar triplets. Such a realization has been scrutinized over the past few decades chiefly in the context of collider studies. In this work we take a complementary approach and investigate whether the model can be probed via the search for a stochastic gravitational wave background induced by the phase transition in which $SU(3)_C \times SU(2)_L \times SU(2)_R \times U(1)_{B-L}$ is broken down to the Standard Model gauge symmetry group. A prerequisite for gravitational wave production in this context is a first-order phase transition, the occurrence of which we find in a significant portion of the parameter space. Although the produced gravitational waves are typically too weak for a discovery at any current or future detector, upon investigating correlations between all relevant terms in the scalar potential, we have identified values of parameters leading to observable signals. This indicates that, given a certain moderate fine-tuning, the minimal left-right symmetric model with scalar triplets features another powerful probe which can lead to either novel constraints or remarkable discoveries in the near future. Let us note that some  of our results, such as the full set of thermal masses, have to the best of our knowledge not been presented before and might be useful for future studies, in particular in the context of electroweak baryogenesis. 
\end{abstract}

\maketitle

\section{Introduction}
\label{sec:intro}
\noindent
The electroweak interaction, associated with the $SU(2)_L\times U(1)_Y$ part of the Standard Model (SM) gauge group, distinguishes left- and right-handed particles and is not symmetric under their interchange. The fundamental principles that would lead to such a low-energy theory are still unknown.
Arguably, one of the most straightforward high-energy extensions of the SM is a realization with restored symmetry between the left and right sectors employing the gauge group $SU(2)_L\times SU(2)_R \times U(1)_{B-L}$ \cite{Pati/Salam,Goran/Mohapatra,Mohapatra/Pati,Mohapatra/Pati2}.
Such a left-right symmetric model (LRSM) is then conventionally broken down spontaneously to the electroweak part of the SM through the presence of additional Higgs degrees of freedom. The pioneering papers on left-right symmetry assumed it to be broken using scalar $SU(2)$ doublets \cite{Goran/Mohapatra,MS}. However, since the discovery of neutrino oscillations it is the scenario with scalar $SU(2)$ triplets \cite{Maiezza:2016ybz,Deshpande:1990ip,Senjanovic:2016bya} that has received significant attention, primarily due to the embedded seesaw mechanism\footnote{Left-right symmetric models with doublets allow for the generation of neutrino mass at tree-level \cite{Brdar:2018sbk} as well as at the quantum level \cite{FileviezPerez:2016erl}; however, an extended scalar or fermion sector is then required.} \cite{Goran,Yanagida:1979as,GellMann:1980vs,Minkowski}. An additional attractive feature of LRSMs as such is the fact that they can be naturally embedded in an $SO(10)$-based grand unified theory~\cite{Fritzsch:1974nn, Georgi:1974my}. Broader studies of $SO(10)$ unification scenarios with left-right symmetric intermediate breaking scale can be found \eg in Refs.~\cite{Arbelaez:2013nga,Deppisch:2017xhv}.

The LRSM incorporating scalar triplets has been widely investigated in the context of present and near-future collider studies, \eg \cite{Chen:2013fna,Lindner:2016lxq,Patra:2015bga,Dev:2015kca,Rodriguez:2002ey,Nemevsek:2018bbt}, which can probe the left-right symmetry breaking scale only up to $\mathcal{O}(\SI{10}{TeV})$. Indeed, the current null results from the LHC exclude the presence of LRSM at such energies, but in order to go beyond, \ie to significantly higher energies, it is necessary to consider a complementary probe, which is the problem addressed in this paper.
Specifically, we study whether the model in question can be successfully tested via the search for a stochastic gravitational wave background arising from a first-order cosmic phase transition \cite{Witten1984,Hogan1984,Hogan1986b,Turner1990a,Kamionkowski1993,
Grojean2007,Ellis:2018mja}.
While in the minimal SM neither the electroweak, nor the QCD phase transition is of first-order \cite{Kajantie1996a,Kajantie1996b,Aoki2006a,Aoki2006b,Bhattacharya2014}, beyond-the-Standard Model (BSM) theories offer such a possibility.
Correspondingly, there exists a large number of studies on gravitational waves induced by cosmic phase transitions in theories addressing various problems of BSM physics, such as neutrino masses \cite{Okada2018a,Brdar:2018num,Dror:2019syi,Hasegawa:2019amx} or dark and hidden sectors, \eg \cite{Schwaller2015,Jaeckel2016a,Dev:2016feu,Baldes2017,Baldes:2018emh,Breitbach:2018ddu,Fairbairn:2019xog,Helmboldt:2019pan,Croon:2018kqn,Mohamadnejad:2019vzg,Dev:2019njv}.
In particular, extended Higgs sectors, such as the one present in LRSMs, can yield a first-order phase transition. One possiblity is that a newly introduced scalar acquires a non-vanishing vacuum expectation value (VEV) in a first-order phase transition.
Another option is that the electroweak phase transition becomes of first order due to the presence of additional terms in the scalar potential.
In this paper we study the former realization, namely we investigate the conditions under which the breaking of $SU(2)_{R}\times U(1)_{B-L}$ occurs via a first-order phase transition.
While we do not consider the latter possibility, we refer the interested reader to Refs.~\cite{Barenboim:1998ib,Choi:1992wb} where the nature of the electroweak phase transition in a LRSM was examined with the purpose of estimating the amount of baryon asymmetry produced through electroweak baryogenesis.
We also wish to note the existence of another study where the phase transition was analyzed and the gravitational wave signature in the left-right symmetric model with triplets was presented for an unconventional symmetry breaking sequence \cite{Sagunski:2012pzo}.
In this work we present a more complete study, in which all the relevant ingredients in the temperature-dependent thermal potential are included and the analysis is independent of the existence of flat directions.

By performing a general scan over the model's parameter space, we find that the left-right breaking phase transition is of first order quite frequently. The amplitude of the produced gravitational wave spectrum then typically turns out to be too small for a successful detection at both space-based and ground-based gravitational wave observatories. However,  more thorough scans, performed after a dedicated investigation of correlations between parameters in the potential, unraveled that in a certain parameter subspace the strength of the gravitational wave signal can enhance drastically, entering the sensitivity region of planned space-based interferometers such as LISA \cite{Caprini:2015zlo}, DECIGO \cite{Seto:2001qf} and BBO \cite{Corbin:2005ny}.

The subsequent text is organized as follows. In \cref{sec:model} we present the considered scenario at tree-level, while in \cref{sec:effV} the temperature dependent effective potential is introduced. \cref{sec:benchmarks} then details the method used for finding viable parameter points and introduces those that are further investigated in the context of phase transition. In \cref{sec:phase-transition} we analyze the phase transition properties with the goal to extract the relevant parameters required for estimating gravitational wave signature. The calculated gravitational wave spectra are presented in \cref{sec:gw}, where we also make a comparison with the sensitivities of the future space-based detectors. Finally, we conclude in \cref{sec:summary}.

\section{The Minimal Left-Right Symmetric Model}
\label{sec:model}
\noindent
The key feature of a minimal LRSM is the restoration of the left-right discrete symmetry at a certain high-energy scale $v_R$ set by the vacuum expectation value (VEV) of a right-handed scalar field.
Hence, the corresponding Lagrangian is left-right symmetric and the $V-A$ structure of the SM physics is only obtained after spontaneous left-right symmetry breaking.
The fermionic particle content of the LRSM can therefore be obtained by a straightforward left-right symmetric extension of the SM content, \ie the right-handed fermion doublets\footnote{Here, the representations are labeled in the usual way in the order $\{SU(3)_C, SU(2)_L, SU(2)_R, U(1)_{B-L}\}$.}, \mbox{$\ell_{R}^{i} \equiv \{\mathbf{1},\mathbf{1},\mathbf{2},-1\}$} and \mbox{$Q^{i}_{R} \equiv \{\mathbf{1},\mathbf{1},\mathbf{2},\frac{1}{3}\}$}, are introduced as
\begin{align}
\ell_{L}^{i} = \begin{pmatrix}
\nu_{L}^{i} \\
e_{L}^{i}
\end{pmatrix}
&\xleftrightarrow{\text{L-R}}
\begin{pmatrix}
\nu_{R}^{i} \\
e_{R}^{i}
\end{pmatrix} = \ell_{R}^{i} \fineq{,}
\\
Q_{L}^{i} = \begin{pmatrix}
u_{L}^{i} \\
d_{L}^{i}
\end{pmatrix}
&\xleftrightarrow{\text{L-R}}
\begin{pmatrix}
u^{i}_{R} \\
d^{i}_{R}
\end{pmatrix} = Q^{i}_{R} \fineq{.}
\end{align}
The presence of the right-handed neutrino partners not only allows for the type-I seesaw mechanism, but it is also essential for cancellation of the $B-L$ gauge anomaly, as can be verified using the left-right symmetric definition of electric charge
\begin{align}\label{eq:LRcharge}
Q = T_{3L} + T_{3R} + \frac{B-L}{2} \fineq{,}
\end{align}
where $T_3$ stands for the third component of weak isospin.

Given the gauge group of LRSMs their gauge sector incorporates a right-handed triplet $W^{\mu\,a}_{R}\equiv\{\mathbf{1},\mathbf{1},\mathbf{3},0\}$, a left-handed triplet $W^{\mu\,a}_{L}\equiv\{\mathbf{1},\mathbf{3},\mathbf{1},0\}$ and a singlet $B^{\mu}\equiv \{\mathbf{1},\mathbf{1},\mathbf{1},0\}$.
The right-handed vector bosons become massive after left-right symmetry breaking, and thus are typically expected to be much heavier than the $SU(2)_L$ vector bosons.

As mentioned above, the Higgs sector of LRSMs can vary.
The minimal breaking scenarios typically include a scalar bi-doublet $\Phi \equiv \{\mathbf{1},\mathbf{2},\mathbf{2},0\}$ incorporating the SM Higgs and a pair of either triplets $\Delta_L \equiv \{\mathbf{1},\mathbf{3},\mathbf{1},-2\},\ \Delta_R \equiv \{\mathbf{1},\mathbf{1},\mathbf{3},-2\}$, or doublets $\chi_L \equiv \{\mathbf{1},\mathbf{2},\mathbf{1},-1\},\ \chi_R \equiv \{\mathbf{1},\mathbf{1},\mathbf{2},-1\}$, which breaks the left-right gauge group down to that of the SM.
In fact, the right-handed scalar (either triplet, or doublet) is enough to do so, but inclusion of its left-handed counterpart ensures that the left-right symmetry is preserved (this is sometimes referred to as ``manifest left-right symmetry"), and thus the $SU(2)_L$ and $SU(2)_R$ gauge couplings are equal: $g_L = g_R \equiv g$.

As reasoned in the introduction, we focus in this work on a minimal LRSM with a Higgs sector containing the triplet scalar representations on top of the bidoublet. Conventionally, they can be represented in the following matrix form
\begin{equation}
\Phi=\left(\begin{array}{cc}
\phi_{1}^{0} & \phi_{1}^{+}\\
\phi_{2}^{-} & \phi_{2}^{0}
\end{array}\right) \fineq{,}\qquad \Delta_{L}=\left(\begin{array}{cc}
\frac{1}{\sqrt{2}}\delta_{L}^{+} & \delta_{L}^{++}\\
\delta_{L}^{0} & -\frac{1}{\sqrt{2}}\delta_{L}^{+}
\end{array}\right) \fineq{,}\qquad \Delta_{R}=\left(\begin{array}{cc}
\frac{1}{\sqrt{2}}\delta_{R}^{+} & \delta_{R}^{++}\\
\delta_{R}^{0} & -\frac{1}{\sqrt{2}}\delta_{R}^{+}
\end{array}\right),\label{eq:x-1}
\end{equation}
so that they transform as
\begin{align}
SU(2)_{L}\times SU(2)_{R}: & \ \ \Phi\rightarrow U_{L}\Phi U_{R}^{\dagger} \fineq{,}\quad \ \Delta_{L}\rightarrow U_{L}\Delta_{L}U_{L}^{\dagger} \fineq{,}\quad \ \Delta_{R}\rightarrow U_{R}\Delta_{R}U_{R}^{\dagger} \fineq{,} \label{eq:x-2}\\
U(1)_{B-L}: & \ \ \Phi\rightarrow\Phi \fineq{,} \thinspace\hspace{8.7ex}\Delta_{L}\rightarrow e^{i2\theta}\Delta_{L} \fineq{,}\hspace{5.4ex}\Delta_{R}\rightarrow e^{i2\theta}\Delta_{R} \fineq{,} \label{eq:x-3}
\end{align}
for general symmetry transformations: $U_{L}\in SU(2)_{L}$, $U_{R}\in SU(2)_{R}$,
and $e^{i\theta}\in U(1)_{B-L}$.

Let us now specify the interactions, which will be important for the investigation of the left-right phase transition within the considered minimal LRSM. The scalar gauge-kinetic terms describing the interaction between scalars and gauge bosons read
\begin{align}
\mathcal{L}_\text{gauge} =	\tr \left[ (D_\mu \Delta_L)^\dagger (D^\mu \Delta_L) \right] + \tr \left[ (D_\mu \Delta_R)^\dagger (D^\mu \Delta_R) \right] + \tr \left[ (D_\mu \Phi)^\dagger (D^\mu \Phi) \right] \fineq{,}
\end{align}
where the gauge-covariant derivatives acting on the scalar multiplets have the usual form
\begin{align}
	D_\mu \Delta & = \del_\mu\Delta - \I g [W^a_\mu \tau^a,\Delta] - \I g_\mathsmaller{B-L} B_\mu \Delta \fineq{,} \\
	D_\mu \Phi & = \del_\mu \Phi - \I g \left( W^a_{L\mu} \tau^a \Phi - \Phi W^a_{R\mu} \tau^a \right) \fineq{.}
\end{align}
Here, $\tau^a=\sigma^a/2$ are the generators of SU(2) and $\sigma^a$ are the Pauli matrices.

Given the scalar sector of our LRSM, one can write down Yukawa terms involving either the bi-doublet, or the triplets. The former option, which reproduces the SM Yukawa terms after the left-right symmetry breaking, is given by
\begin{align}
	\mathcal{L}_\text{Yukawa} \supseteq \overline{Q}_{Li} (y_{ij}\Phi + \tilde{y}_{ij}\tilde{\Phi}) Q_{Rj} +\hc\fineq{,}
\end{align}
where $\tilde{\Phi}\equiv\sigma_2\Phi^*\sigma_2$ and $i,j$ are quark generation indices running from 1 to 3. Since we always consider the case $\kappa_2\ll\kappa_1\simeq v_\tinytext{EW}$, it is sufficient to assume
$y=\operatorname{diag}(0,0,y_t)$ and $\tilde{y}=0$.
The Lagrangian then reduces to
\begin{align}
	\mathcal{L}_\text{Yukawa} \supseteq y_t ( \overline{Q}_{L3}\Phi Q_{R3} + \overline{Q}_{R3}\Phi^\dagger Q_{L3} ) \fineq{,}
\label{eq:yukDoublet}
\end{align}
where $Q_{L3}=(t_L, b_L)^\transpose$ and $Q_{R3}=(t_R, b_R)^\transpose$.

The Yukawa terms involving the scalar triplets read
\begin{align}
	\mathcal{L}_\text{Yukawa} \supseteq -Y_M^{ij} \left[ \overline{\ell_L^c}_i \I\sigma_2\Delta_L \ell_{Lj} + (L\leftrightarrow R) \right] + \hc \fineq{,}
\end{align}
where we again sum over all three generations. For the sake of simplicity, we will assume that the matrix of Majorana Yukawa couplings has the structure
\begin{align}
	Y_M^{ij} = y_M \delta^{ij} \fineq{.}
\end{align}
with an $\mathcal{O}(1)$ real coupling $y_M$.
Hence, it is sufficient to investigate the interactions for one fermion generation and then multiply all diagrams containing lepton loops with the number of generations $N_g=3$.
Also, since the terms involving the left- and the right-handed triplet are structurally identical, it suffices to concentrate on the left sector.
In summary, we consider the terms
\begin{align}
	\mathcal{L}_\text{Yukawa} \supseteq -y_M \overline{\ell_L^c} \I\sigma_2\Delta_L \ell_{L}
	+ y_M \overline{\ell_L} \Delta_L^\dagger\I\sigma_2 \ell^c_L + (L\leftrightarrow R) \fineq{.}
\label{eq:yukTriplet}
\end{align}

The full scalar tree-level potential including all renormalizable terms that can be built out of the given particle content and that are allowed by the gauge symmetry reads
\begin{align}
V_\text{tree} = V_{\Phi} + V_{\Delta} + V_{\Phi\Delta} \fineq{,}
\end{align}
where
\begin{align}
V_{\Phi} =\ &  -\mu_{1}^{2}\text{Tr}[\Phi^{\dagger}\Phi]-\mu_{2}^{2}(\text{Tr}[\tilde{\Phi}\Phi^{\dagger}]+\text{Tr}[\tilde{\Phi}^{\dagger}\Phi])-\mu_{3}^{2}(\text{Tr}[\Delta_{L}\Delta_{L}^{\dagger}]+\text{Tr}[\Delta_{R}\Delta_{R}^{\dagger}])+\lambda_{1}\text{Tr}[\Phi^{\dagger}\Phi]^{2} \nonumber \\
 & +\lambda_{2}\left(\text{Tr}[\tilde{\Phi}\Phi^{\dagger}]^{2}+\text{Tr}[\tilde{\Phi}^{\dagger}\Phi]^{2}\right)+\lambda_{3}\text{Tr}[\tilde{\Phi}\Phi^{\dagger}]\text{Tr}[\tilde{\Phi}^{\dagger}\Phi]+\lambda_{4}\text{Tr}[\Phi^{\dagger}\Phi](\text{Tr}[\tilde{\Phi}\Phi^{\dagger}]+\text{Tr}[\tilde{\Phi}^{\dagger}\Phi]) \fineq{,} \nonumber \\
V_{\Delta} =\ & \rho_{1}\left(\text{Tr}[\Delta_{L}\Delta_{L}^{\dagger}]^{2}+\text{Tr}[\Delta_{R}\Delta_{R}^{\dagger}]^{2}\right)+\rho_{2}(\text{Tr}[\Delta_{L}\Delta_{L}]\text{Tr}[\Delta_{L}^{\dagger}\Delta_{L}^{\dagger}]+\text{Tr}[\Delta_{R}\Delta_{R}]\text{Tr}[\Delta_{R}^{\dagger}\Delta_{R}^{\dagger}])\nonumber \\
 & +\rho_{3}\text{Tr}[\Delta_{L}\Delta_{L}^{\dagger}]\text{Tr}[\Delta_{R}\Delta_{R}^{\dagger}]+\rho_{4}(\text{Tr}[\Delta_{L}\Delta_{L}]\text{Tr}[\Delta_{R}^{\dagger}\Delta_{R}^{\dagger}]+\text{Tr}[\Delta_{L}^{\dagger}\Delta_{L}^{\dagger}]\text{Tr}[\Delta_{R}\Delta_{R}]) \fineq{,} \nonumber \\
V_{\Phi\Delta} =\ & \alpha_{1}\text{Tr}[\Phi^{\dagger}\Phi](\text{Tr}[\Delta_{L}\Delta_{L}^{\dagger}]+\text{Tr}[\Delta_{R}\Delta_{R}^{\dagger}])+\alpha_{3}(\text{Tr}[\Phi\Phi^{\dagger}\Delta_{L}\Delta_{L}^{\dagger}]+\text{Tr}[\Phi^{\dagger}\Phi\Delta_{R}\Delta_{R}^{\dagger}])\nonumber \\
 & +\alpha_{2}(\text{Tr}[\Delta_{L}\Delta_{L}^{\dagger}]\text{Tr}[\tilde{\Phi}\Phi^{\dagger}]+\text{Tr}[\Delta_{R}\Delta_{R}^{\dagger}]\text{Tr}[\tilde{\Phi}^{\dagger}\Phi]+{\rm h.c.})\nonumber \\
 & +\beta_{1}(\text{Tr}[\Phi\Delta_{R}\Phi^{\dagger}\Delta_{L}^{\dagger}]+\text{Tr}[\Phi^{\dagger}\Delta_{L}\Phi\Delta_{R}^{\dagger}])+\beta_{2}(\text{Tr}[\tilde{\Phi}\Delta_{R}\Phi^{\dagger}\Delta_{L}^{\dagger}]+\text{Tr}[\tilde{\Phi}^{\dagger}\Delta_{L}\Phi\Delta_{R}^{\dagger}])\nonumber \\
 & +\beta_{3}(\text{Tr}[\Phi\Delta_{R}\text{\ensuremath{\tilde{\Phi}^{\dagger}\Delta_{L}^{\dagger}}}]+\text{Tr}[\Phi^{\dagger}\Delta_{L}\text{\ensuremath{\tilde{\Phi}\Delta_{R}^{\dagger}}}]) \fineq{,} \label{eq:V}
\end{align}
with $\tilde{\Phi}\equiv\sigma_{2}\Phi^{*}\sigma_{2}$. All the couplings are assumed to be real for simplicity\footnote{In principle, some couplings such as $\alpha_{2}$ and $\lambda_{4}$ could be complex. However, this depends on whether the ${\cal C}$-parity
or the ${\cal P}$-parity is introduced in the model. The assumption of real couplings can be compatible with both scenarios. For more details, see Eq.~(9) and Eq.~(10) in Ref.~\cite{Maiezza:2016ybz}, where the differences  have been addressed.}. The Higgses appearing in a viable LRSM scalar potential are expected to acquire the following vacuum expectation values (VEVs)
\begin{equation}
\langle\Phi\rangle=\frac{1}{\sqrt{2}}\left(\begin{array}{cc}
\kappa_{1} & 0\\
0 & \kappa_{2}e^{i\theta_{2}}
\end{array}\right) \fineq{,}\qquad \langle\Delta_{L}\rangle=\frac{1}{\sqrt{2}}\left(\begin{array}{cc}
0 & 0\\
v_{L}e^{i\theta_{L}} & 0
\end{array}\right) \fineq{,}\qquad \langle\Delta_{R}\rangle=\frac{1}{\sqrt{2}}\left(\begin{array}{cc}
0 & 0\\
v_{R} & 0
\end{array}\right) \fineq{.} \label{eq:x-5}
\end{equation}
However, this is not guaranteed for general values of the couplings in Eq.~(\ref{eq:V}). As it has been studied in 
Refs.~\cite{Dev:2018foq,Chauhan:2019fji}, only a part of the full parameter space is able to get such VEVs. We will adopt the numerical method of Ref.~\cite{Dev:2018foq} to identify the part of parameter space where Eq.~(\ref{eq:x-5}) is a global minimum of the potential. The technical details will be explained in \cref{sec:benchmarks}.

In the LRSM, the bidoublet VEVs are expected to be at the electroweak scale
\begin{equation}
\sqrt{\kappa_{1}^{2}+\kappa_{2}^{2}}=v\approx246\ {\rm GeV} \fineq{.} \label{eq:x-4-1}
\end{equation}
For later use, we also introduce the $\tan\beta$ parameter, defined as 
\begin{equation}
\tan\beta=\frac{\kappa_{2}}{\kappa_{1}} \fineq{.} \label{eq:x-16}
\end{equation}
The triplet VEVs should be either much higher ($v_{R}$) or much lower ($v_{L}$) than the the electroweak scale.  They are connected to $\kappa_{1}$ and $\kappa_{2}$ by the well-known seesaw relation
of VEVs \cite{Deshpande:1990ip}
\begin{equation}
\beta_{1}\kappa_{1}\kappa_{2}\cos\left(\theta_{2} - \theta_{L}\right) + \beta_{2}\kappa_{1}^{2}\cos\theta_{L} + \beta_{3}\kappa_{2}^{2}\cos\left(2\theta_{2} - \theta_{L}\right) = (2\rho_{1} - \rho_{3})v_{L}v_{R} \fineq{,} \label{eq:x-6}
\end{equation}
which can be derived from the equations of vanishing first-order derivatives.
Therefore, if $\beta_{1}$, $\beta_{2}$, and $\beta_{3}$ are set to zero, then $v_{L}$ will be zero and $v_{R}$ can be arbitrarily high.
In this case, light neutrino masses are generated only by type-I seesaw.
For simplicity, throughout this paper we always keep $\beta_{1}=\beta_{2}=\beta_{3}=0$, and consequently $v_{L}=0$. In addition, we also set $\theta_{2}=0$, which can be justified using the conclusion that spontaneous CP symmetry breaking does not appear for a considerably large part of the parameter space according to the numerical study in \cite{Dev:2018foq}. 

In summary, in this paper with the assumptions that all the potential parameters are real and $\beta_{1}=\beta_{2}=\beta_{3}=0$, we only consider the following viable VEV alignments: 
\begin{equation}
\langle\Phi\rangle=\frac{1}{\sqrt{2}}\left(\begin{array}{cc}
\kappa_{1} & 0\\
0 & \kappa_{2}
\end{array}\right) \fineq{,}\qquad \langle\Delta_{L}\rangle=0\fineq{,}\qquad \langle\Delta_{R}\rangle=\frac{1}{\sqrt{2}}\left(\begin{array}{cc}
0 & 0\\
v_{R} & 0
\end{array}\right) \fineq{.}\label{eq:x-5-1}
\end{equation}
Then the potential at the minimum is
\begin{eqnarray}
V_\text{min} & = & -\frac{1}{2}\mu_{1}^{2}\left(\kappa_{1}^{2}+\kappa_{2}^{2}\right)-2\mu_{2}^{2}\kappa_{1}\kappa_{2}-\frac{1}{2}\mu_{3}^{2}v_{R}^{2}\nonumber \\
 &  & +\frac{1}{4}\text{ }\lambda_{1}\left(\kappa_{1}^{2}+\kappa_{2}^{2}\right){}^{2}+2\lambda_{2}\kappa_{1}^{2}\kappa_{2}^{2}+\lambda_{3}\kappa_{1}^{2}\kappa_{2}^{2}+\lambda_{4}\left(\kappa_{1}^{3}\kappa_{2}+\kappa_{1}\kappa_{2}^{3}\right)+\frac{1}{4}v_{R}^{4}\rho_{1}\nonumber \\
 &  & +\frac{1}{4}\alpha_{1}\left(v_{R}^{2}\kappa_{1}^{2}+v_{R}^{2}\kappa_{2}^{2}\right)+\alpha_{2}v_{R}^{2}\text{ }\kappa_{1}\kappa_{2}+\frac{1}{4}\alpha_{3}v_{R}^{2}\kappa_{2}^{2} \fineq{.} \label{eq:V0}
\end{eqnarray}
From $\partial V_\text{min}/\partial\kappa_{1}=\partial V_\text{min}/\partial\kappa_{2}=\partial V_\text{min}/\partial v_{R}=0$,
one can obtain
\begin{subequations}
\begin{eqnarray}
\mu_{1}^{2} & = & \lambda_{1}\left(\kappa_{1}^{2}+\kappa_{2}^{2}\right)+2\kappa_{1}\kappa_{2}\lambda_{4}+\frac{1}{2}v_{R}^{2}\alpha_{1}-\frac{\alpha_{3}}{2}\frac{v_{R}^{2}\text{ }\kappa_{2}^{2}}{\kappa_{1}^{2}-\kappa_{2}^{2}} \fineq{,}\label{eq:x-7}\\
\mu_{2}^{2} & = & \left(2\lambda_{2}+\lambda_{3}\right)\kappa_{1}\kappa_{2}+\frac{\lambda_{4}}{2}\left(\kappa_{1}^{2}+\kappa_{2}^{2}\right)+\frac{\alpha_{2}}{2}v_{R}^{2}+\frac{\alpha_{3}}{4}\frac{v_{R}^{2}\kappa_{1}\kappa_{2}}{\kappa_{1}^{2}-\kappa_{2}^{2}} \fineq{,}\label{eq:x-8}\\
\mu_{3}^{2} & = & \rho_{1}v_{R}^{2}+\frac{1}{2}\alpha_{1}\left(\kappa_{1}^{2}+\kappa_{2}^{2}\right)+2\alpha_{2}\kappa_{1}\kappa_{2}+\frac{1}{2}\alpha_{3}\kappa_{2}^{2} \fineq{,}\label{eq:x-9}
\end{eqnarray}
\end{subequations}
which will be used to determine the quadratic couplings from the VEVs and quartic couplings.

As we concentrate on LRSMs with a left-right symmetry breaking scale well beyond the reach of current collider searches, we will assume in the following that \mbox{$v_R \gg \kappa_1,\kappa_2$}.\footnote{A relatively high scale of $v_{R}$ is important for the new particles
to have heavy masses above the LHC bounds but this may be not sufficient.
 Eventually for the specific scenarios discussed in this work, we
further check the full mass spectrum of new particles to make sure
that they are compatible with various collider searches.}
Hence, we will be primarily interested in evaluating the effective potential in regions of scalar field space, where the neutral component $\delta_R^0$ of the right-handed triplet attains much larger absolute values than the remaining neutral fields $\phi_1^0$, $\phi_2^0$ and $\delta_L^0$.
It is therefore well justified to approximately regard the effective potential as a function of only a single field, namely of the real part of $\delta_r^0$.
Specifically, the tree-level potential in Eq.~\eqref{eq:V0} can then be simplified to the expression
\begin{align}
V_{0}(\dr) = -\tfrac{1}{2}\mu_{3}^{2}\dr^{2}+\tfrac{1}{4}\rho_{1} \dr^{4}
\quad\text{with}\quad
r := \operatorname{Re}{\delta_R^0}/\sqrt{2} \fineq{,}
\label{eq:V0simple}
\end{align}
which is used for our further calculations.

\section{Finite-Temperature Effective Potential}
\label{sec:effV}
\noindent
In order to study the left-right phase transition we need to go to the quantum level and construct the effective potential corresponding to the above defined model. The one-loop daisy-improved finite-temperature effective potential as a function of $\dr$ and temperature $T$ can be written as
\begin{align}
V_{\text{eff}}(\dr,T) = V_{0}(\dr) + V_{\text{CW}}(\dr) + V_{\text{FT}}(\dr,T) + V_{\text{D}}(\dr,T) \fineq{.}
\label{eq:Veff}
\end{align}
Here, $V_{0}(\dr)$ is the tree-level potential from Eq.~\eqref{eq:V0simple}.
The temperature-independent Coleman-Weinberg effective potential $V_{\text{CW}}$ in the $\MSbar$ scheme and the Landau gauge is given by
\begin{align}
 V_{\text{CW}}(\dr) 
 &= \frac{1}{64\pi^2}\bigg[\sum_{i}m_i^4(\dr)\bigg(\log\frac{m_i^2(\dr)}{\mu^2}-\frac{3}{2}\bigg)
 + 6 m_{W_R}^4(\dr)\bigg(\log\frac{m_{W_R}^2(\dr)}{\mu^2}-\frac{5}{6}\bigg) \nn \\
 &\phantom{= \frac{1}{64\pi^2}\bigg[}+ 3 m_{Z_R}^4(\dr)\bigg(\log\frac{m_{Z_R}^2(\dr)}{\mu^2}-\frac{5}{6} \bigg)
 - 6 m_{\nu_R}^4(\dr)\bigg(\log\frac{m_{\nu_R}^2(\dr)}{\mu^2}-\frac{3}{2}
\bigg)\bigg] \fineq{,}
\end{align}
where the sum in the first term in the square brackets runs over the scalar spectrum of our model, while the other terms correspond to the gauge bosons and the right-handed neutrino. All the field-dependent tree-level masses $m_i$ are given in Appendix~\ref{app:spectra}. The $\MSbar$ renormalization scale $\mu$ is in our later calculations set to the value of $v_R$.

The thermal effective potential $V_{\text{FT}}$ reads
\begin{align}
V_{\text{FT}}(\dr,T) &= \frac{T^4}{2\pi^2}\left[
\sum_{i}J_{\text{B}}\bigg(\frac{m^2_i(\dr)}{T^2}\bigg)
+6J_{\text{B}}\bigg(\frac{m^2_{W_R}(\dr)}{T^2}\bigg)
+3J_{\text{B}}\bigg(\frac{m^2_{Z_R}(\dr)}{T^2}\bigg)
-6J_{\text{F}}\bigg(\frac{m^2_{\nu_R}(\dr)}{T^2}\bigg)\right] \fineq{.}
\end{align}
Here, the thermal functions $J_{\text{B}}$ and $J_{\text{F}}$ for bosons and fermions, respectively, are defined as
\begin{align}
J_{\text{B/F}}(r^2) = \int_0^\infty \mathrm{d}x\, x^2 \log\left(1\pm \mathrm{e}^{\sqrt{x^2+r^2}}\right).
\end{align}
The last term in Eq.~\ref{eq:Veff} stands for the resummed daisy diagrams representing the leading infrared divergent higher-loop contributions. This part of the effective potential is given by
\begin{align}
V_{\text{D}}(\dr,T) = -\frac{T}{12\pi}\sum_{i}\bigg[M_i^3(\dr)-m_i^3(\dr)\bigg],
\label{eq:VD}
\end{align}
where $M_i^2(\dr)$ are the thermal masses obtained as the eigenvalues of the matrix \mbox{$\mathcal{M}_i^2(\dr)+\Pi_i(\dr,T)$} with $\mathcal{M}_i^2(\dr)$ being the tree-level mass matrices discussed in Appendix~\ref{app:spectra} and $\Pi_i(\dr,T)$ standing for the matrices of thermal self-energies provided explicitly in Appendix~\ref{app:thermalmasses}. The sum runs over all the bosons present in the studied LRSM.

\section{Generating Successful Scenarios}
\label{sec:benchmarks}
\noindent
To facilitate the later study of phase transitions, it is necessary
to create some benchmarks with all the potential parameters numerically
given. We shall choose the values of potential parameters in such
a way that both theoretical and phenomenological requirements
are satisfied, including:
\begin{enumerate}
\item The potential at the tree level is bounded from below (BFB);
\item The potential has a global minimum with the predefined VEVs;
\item The VEVs satisfy: $v\approx246\ {\rm GeV}$ and $v_{R}\gtrsim10^{4}$
GeV (to make $W_{R}^{\pm}$ sufficiently heavy and thus satisfy the LHC bounds);
\item The physical spectrum contains a scalar with mass \mbox{$m_{h}\approx125\ {\rm GeV}$} and the properties of the SM Higgs boson; all the
other bosons (except for the six Goldstone bosons) have masses at
the same order as $v_{R}$. 
\end{enumerate}

The first two requirements are purely theoretical. So far there has 
been no straightforward analytical procedure that can be used to thoroughly infer whether
a given sufficiently complicated potential can fully satisfy the BFB and global minimum requirements. Thus we need to adopt a numerical method (which will be described shortly afterwards)
to check the BFB condition and to search
for global minima of the potential. During random generation of
numerical samples, to improve the chance of obtaining BFB potentials,
we assume all quartic couplings are non-negative.

The third requirement can be met by taking $v$, $v_{R}$, and $\tan\beta$
as input parameters and using Eqs.~(\ref{eq:x-7}), (\ref{eq:x-8}), and (\ref{eq:x-9}) to determine $\mu_{1}^{2}$, $\mu_{2}^{2}$ and $\mu_{3}^{2}$. The last requirement can be simplified if $\alpha_{1}=0$ and
$\tan\beta$ is small. In this limit, the full SM Higgs mass
\begin{align}
m_{h}^{2} =\ & \frac{v^{2}}{2\rho_{1}}\left[4\lambda_{1}\rho_{1}-\alpha_{1}^{2}+c_{\beta}s_{\beta}\left(-8\alpha_{1}\alpha_{2}+16\lambda_{4}\rho_{1}\right)\right.\nonumber \\
 & -2s_{\beta}^{2}\left(8\alpha_{2}^{2}+\alpha_{1}\alpha_{3}-8\left(2\lambda_{2}+\lambda_{3}\right)\rho_{1}\right)-8c_{\beta}s_{\beta}^{3}\alpha_{2}\alpha_{3}\nonumber \\
 & \left.+s_{\beta}^{4}\left(16\alpha_{2}^{2}-\alpha_{3}^{2}-16\left(2\lambda_{2}+\lambda_{3}\right)\rho_{1}\right)\right]+{\cal O}(v^{4}) \label{eq:x-15}
\end{align}
can be approximated as $m_{h}^{2}\approx2\lambda_{1}v^{2}$, which requires
$\lambda_{1}\approx\frac{1}{2}m_{h}^{2}/v^{2}\approx0.13$. 

Combining the above requirements together,  we scan the following parameter space:
\begin{subequations}
\begin{align}
v & =246\ {\rm GeV},\ v_{R}\in[10^4,\ 10^6]\, {\rm GeV},\ \tan\beta=\tan10^{-3}, \label{eq:para_space_v}\\
\lambda_{1} & =0.13,\ \lambda_{2}=0,\ \lambda_{3}\in[0,\ 2],\ \lambda_{4}=0, \label{eq:para_space_l}\\
\rho_{1} & \in[0,\ 0.5],\ \rho_{2}\in[0,\ 2],\ \rho_{3}\in[1,\ 2],\ \rho_{4}=0, \label{eq:para_space_r}\\
\alpha_{1} & =0,\ \alpha_{2}\in[0,\ 0.5],\ \alpha_{3}\in[0,\ 1],  \label{eq:para_space_a}\\
\beta_{1}&=\beta_{2}=\beta_{3}=0.
\label{eq:para_space_b}
\end{align}
\label{eq:para_space}%
\end{subequations}
Here some quartic couplings are set to zero to simplify the potential and the analysis.
Note that we should keep sufficiently many quartic couplings nonzero to meet
the four requirements mentioned above. For example, $\alpha_{i}$
and $\beta_{i}$ ($i=1$, 2, 3) cannot be zero simultaneously, otherwise
the potential would have more massless eigenstates.  In addition,
$\rho_{1}$ and $\rho_{3}$ in Eq.~(\ref{eq:para_space_r}) are set in such
a way that $\rho_{3}$ is always larger than $2\rho_{1}$, which
increases the
probability of obtaining successful samples in random generation.

Within the parameter space specified by Eqs.~(\ref{eq:para_space_v}) to (\ref{eq:para_space_b}), we randomly generate 100 samples. Because for each sample all the potential parameters in Eq.~(\ref{eq:V})
are numerically given or determined, we can numerically minimize
each scalar potential. In this work, we adopt {\tt Mathematica}'s built-in function {\tt NMinimize} for numerical minimization.
During the numerical process, if the potential is
not BFB, then the iterative processes of numerical minimization will
be divergent, which can be obviously detected.

\def\arraystretch{1.3}
\begin{table*}
\begin{ruledtabular}
\begin{tabular}{ccccc}
			  & BP1 & BP2 & BP3 & BP4 \tabularnewline
\cline{2-5} 
$v/{\rm GeV}$ & 246 & 246 & 246 & 246\tabularnewline
$v_{R}/{\rm GeV}$ & $10^{4}$ & $10^{6}$ & $10^{4}$ & $5\times 10^{4}$\tabularnewline
$\tan\beta$ & $10^{-3}$ & $10^{-3}$ & 0 & 0\tabularnewline
$\lambda_{1}$ & 0.13 & 0.13 & 0.13 & 0.13\tabularnewline
$\lambda_{2}$ & 0 & 0 & 0 & 0\tabularnewline
$\lambda_{3}$ & 1.2040 & 0.88814 & 0.6 & 0.6\tabularnewline
$\lambda_{4}$ & 0 & 0 & 0 & 0\tabularnewline
$\rho_{1}$ & 0.13414 & 0.11146 & 0.001 & 0.002\tabularnewline
$\rho_{2}$ & 1.2613 & 1.4109 & 0.900218 & 0.401126\tabularnewline
$\rho_{3}$ & 1.5140 & 1.5489 & 0.900215 & 0.401126\tabularnewline
$\rho_{4}$ & 0 & 0 & 0 & 0.040113\tabularnewline
$\alpha_{1}$ & 0 & 0 & 0 & 0\tabularnewline
$\alpha_{2}$ & 0.30246 & 0.15557 & 0 & 0\tabularnewline
$\alpha_{3}$ & 0.10765 & 0.11185 & 1.14815 & 0.378138\tabularnewline
$\beta_{1,\thinspace2,\thinspace3}$ & 0 & 0 & 0 & 0\tabularnewline
$g$ & 0.65 & 0.65 & 0.65 & 0.65\tabularnewline
$g_{B-L}$ & 0.4324 & 0.4324 & 0.4324 & 0.4324\tabularnewline
$y_t$ & 0.95 & 0.95 & 0.95 & 0.95\tabularnewline
$y_{M}$ & 1 & 1 & 0.78595  & 0.52404 \tabularnewline
\end{tabular}
\end{ruledtabular}
\caption{\label{tab:bench} Numerical values of the four selected benchmarks studied in this work.}
\end{table*}

To find the global minimum of the potential, we repeat the minimization
several times, each time with a different initial searching point
for the iterative processes. This can be achieved by setting the random
seed of {\tt NMinimize}. Among the minima obtained in the repetition,
the deepest one is  expected to be the global minimum at a high confidence
level. It should be noticed that due to some discrete symmetries
(e.g., the parity symmetry) the potential always has multiple global
minima with an equal depth. So in general, if a global minimum is
obtained in this way, it may be of a different form as Eq.~(\ref{eq:x-5-1}).
To solve this problem, we also minimize the potential $V_{0}$ in
Eq.~(\ref{eq:V0}), with respect to $\kappa_{1}$, $\kappa_{2}$ and
$v_{R}$. If the minimum obtained in such way has the same depth as
the global minimum obtained in the general minimization process, then
we label this sample as a viable benchmark.


After the above numerical process, we obtain 74 viable samples from the 100 random samples. To understand the effect of higher values of $v_R$, we also generate another set of samples with the same parameter setting as Eqs.~(\ref{eq:para_space_v}) to (\ref{eq:para_space_b}) except for $v_R=10^6$ GeV. In this case, we obtain 82 viable samples. All these samples are  further passed to the next step  for the study of phase transition and GW signals. Two of them are selected as benchmarks (BP1 and BP2) in our study, and their numerical values are listed in Tab.~\ref{tab:bench}.

In Sec.~\ref{sec:phase-transition}, we will demonstrate that in order to have significant GW signals, the potential should have sufficiently small $\rho_1$. Hinted by the correlation of the GW signals and $\rho_1$, we shall inspect some cases with small $\rho_1$. When $\rho_1$ is very small, however, the $\rho_1 v_R^4$ term would be subdominant, leading to a more complicated scenario. For simplicity, we would like to maintain the $\rho_1 v_R^4$ dominance in the potential. Hence we choose $\rho_1=10^{-3}$ or $2\times 10^{-3}$ so that $\rho_1 v_R^4$ should be dominant over the $\lambda_1$ term. In addition, we set $\tan \beta=0$ and $\alpha_1=0$ so that all the $\alpha$ terms do not contribute to $V_0$. 

We should note that a tree-level shallow potential is usually vulnerable to the Coleman-Weinberg correction. For instance, when the potential at the tree level leads to correct values of $v$ and $m_h$, including the Coleman-Weinberg correction may drastically change these values. Technically, one may consider two opposite solutions: the true VEVs are dominantly determined by the Coleman-Weinberg potential, or they are dominantly determined by the tree-level potential. Since the Coleman-Weinberg potential and the tree-level potential have some free parameters to tune, both of the solutions can be achieved. The former would imply that the left-right gauge symmetry is broken radiatively, which could be more involved. For simplicity, we adopt the latter, which means the Coleman-Weinberg potential is tuned to be subdominant even though the tree-level potential is already quite shallow. Such suppression of radiative contribution was already employed in \cite{Nemevsek:2016enw} and here we also make use of the Yukawa coupling to achieve tree-level dominance. In that spirit, we generated another two benchmarks (BP3 and BP4), also listed in Tab.~\ref{tab:bench}, for which $\rho_1$ is much smaller than for BP1 and BP2 but the $\rho_1 v_R^4$ dominance still holds and the Coleman-Weinberg term is subdominant.

\section{Left-Right Phase Transition}
\label{sec:phase-transition}
\noindent
Having thoroughly discussed zero-and finite-temperature aspects of the investigated model in the previous sections, we are now interested in the question of how a parity-breaking vacuum, \mbox{$v_R\neq v_L=0$}, can spontaneously emerge from a symmetric high-temperature groundstate, $v_R=v_L=0$, in the early universe.
The following section will therefore be devoted to an analysis of the corresponding left-right-symmetry-breaking thermal phase transition.

\begin{figure}[t]
	\centering
	\includegraphics[scale=1.0]{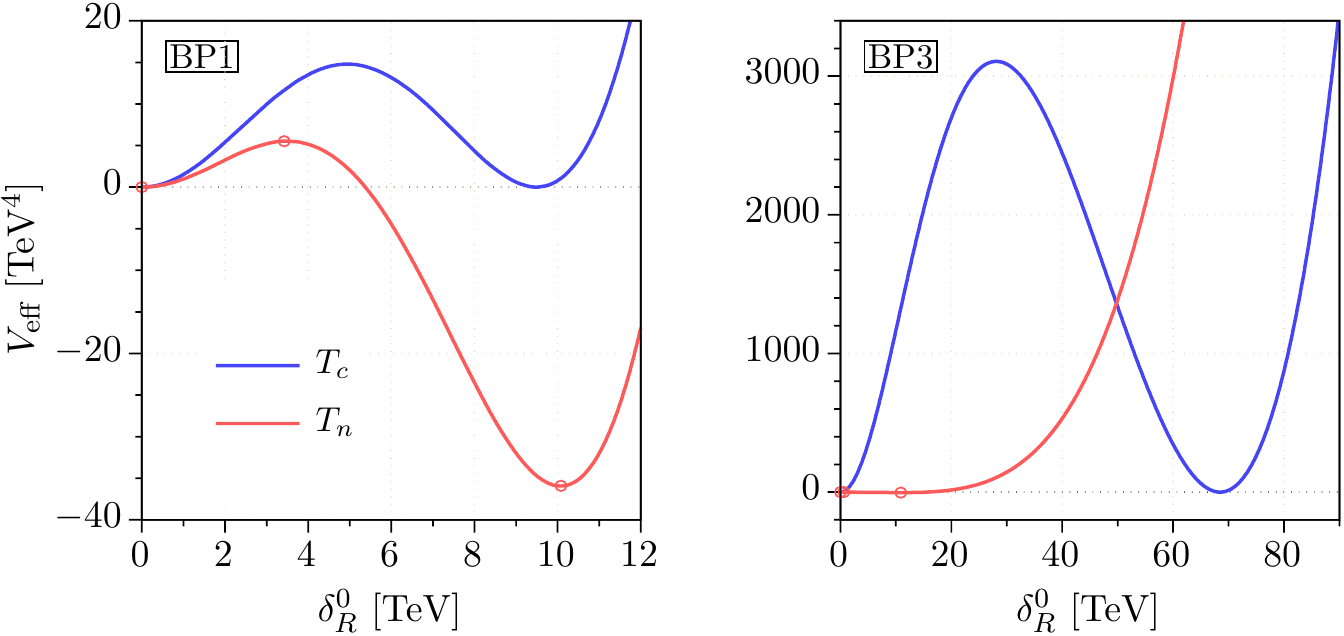}
	\caption{Finite-temperature effective potential of \cref{eq:Veff} for two of the benchmarks of \cref{tab:bench}. For each parameter point we compare the potential at the critical temperature $T_c$ (blue line) with that at the nucleation temperature \mbox{$T_n<T_c$} (red line). The red circles indicate extrema of the potential. The existence of two degenerate local minima at $T_c$ implies that the LR-breaking phase transition is of first order for the considered points. For the actual values of $T_n$ and $T_c$, we refer the reader to \cref{tab:paraGW} in \cref{sec:gw}.}
	\label{fig:pt:potential}
\end{figure}

In order to explore the theory's phase structure, we use the finite-temperature effective potential $V_\text{eff}$ which was introduced in \cref{eq:Veff} of \cref{sec:effV}.
Importantly, the \textit{global} minimum of $V_\text{eff}$ determines the model's true groundstate $v_R(T)$ for a  given temperature $T$.
In particular, following $v_R(T)$ from the left-right (LR) symmetric high-temperature phase to the parity-broken phase at low temperatures allows us to distinguish first- and second-order transitions based on whether or not two degenerate local minima appear at a certain critical temperature $T_c$, see also \cref{fig:pt:potential}.
Our study of the multi-dimensional parameter space defined in \cref{eq:para_space} reveals that both types of transitions generally exist.
 
As we are primarily interested in a possible gravitational wave signature from left-right symmetry breaking, we will in the following concentrate on scenarios where the associated phase transition is of first order.
Such transitions are known to proceed via the nucleation of bubbles within which the scalar fields have already attained the values of the true groundstate (\textit{here}: \mbox{$v_R\neq0$}).
Those bubbles then grow inside an expanding universe that is still in the metastable phase (\textit{here}: \mbox{$v_R=0$}).
The properties of a first-order transition are thus predominantly determined by two temperature-dependent quantities: the bubble nucleation rate $\Gamma$, on the one hand, and the Hubble parameter $H$, on the other hand.
Following Refs.~\cite{Witten1984,Hogan1984,Linde1981}, we estimate the former as
\begin{align}
	\Gamma(T) \simeq T^4 \left( \frac{S_3}{2\pi T} \right)^{\!\frac{3}{2}} \E^{-S_3/T} \fineq{.}
	\label{eq:pt:gamma}
\end{align}
Here, the three-dimensional Euclidean action $S_3$ is to be understood as having been evaluated for the $O(3)$-symmetric tunneling or bounce solution, which, in turn, is obtained by solving the scalar field's equation of motion,
\begin{align}
	\frac{\dd^2 \dr}{\dd x^2} + \frac{2}{x} \frac{\dd \dr}{\dd x} = \frac{\dd V_\text{eff}(\dr, T)}{\dd \dr} \fineq{,}
	\label{eq:pt:eom}
\end{align}
subject to the boundary conditions \mbox{$\dd \dr/\dd x = 0$} at \mbox{$x=0$} and \mbox{$\dr \to 0$} as \mbox{$x\to \infty$} with $x$ denoting the three-dimensional radial coordinate.
In the further course of the present work, we will employ the \texttt{CosmoTransitions} code \cite{Wainwright2012} both to solve \cref{eq:pt:eom} and to compute the resulting Euclidean action $S_3$.

Next, the Hubble parameter is given via Friedmann's equation and can be expressed in terms of the universe's radiation and vacuum energy densities $\rho_\text{rad}$ and $\rho_\text{vac}$, respectively
\begin{align}
	H(T)^2 = \frac{\rho_\text{rad}(T) + \rho_\text{vac}(T)}{3\MPl^2}
	= \frac{1}{3\MPl^2} \left( \frac{\pi^2}{30}g_* T^4 + \Delta V(T) \right) \fineq{.}
	\label{eq:pt:hubble}
\end{align}
In the above equation, \mbox{$g_*=\num{134}$} denotes the effective number of relativistic degrees of freedom in the left-right symmetric model under investigation.
Besides, \mbox{$\MPl=\SI{2.435e18}{GeV}$} is the reduced Planck mass.
The vacuum energy density is calculated as the potential difference between the \textit{local} minimum at \mbox{$\dr=0$} and the \textit{global} one at \mbox{$\dr=v_R(T)$}, \ie \mbox{$\Delta V(T) := V_\text{eff}(0,T)-V_\text{eff}(v_R(T),T)$}.
Note that the vacuum contribution to the $H$ is only expected to be relevant in the case of phase transitions with a considerable amount of supercooling.

Once both the bubble nucleation rate and the Hubble parameter are known, it is straightforward to compute the so-called nucleation temperature $T_n$.
It is defined as the temperature where \textit{one} bubble per horizon volume is created on average, namely
\begin{align}
	\int_{T_n}^{T_c} \frac{\dd T}{T} \frac{\Gamma(T)}{H(T)^4} \stackrel{!}{=} 1 \fineq{.}
	\label{eq:pt:Tn}
\end{align}
As such, $T_n$ is a measure for the temperature at which the phase transition actually occurs and thus crucially influences the associated gravitational wave spectrum, in particular the position of its peak frequency (\cf \cref{eq:fSW,eq:fTURB} in the next section).

Given the nucleation temperature, we can now go on to determine two further phenomenologically important parameters.
First, a measure for the phase transition's strength is provided by the energy released during the transition normalized to the universe's radiation energy density, more precisely (see \eg \cite{Espinosa:2010hh})
\begin{align}
	\alpha = \frac{1}{\rho_\text{rad}(T_n)} \left( \Delta V(T_n) - \frac{T_n}{4} \left. \frac{\del \Delta V(T)}{\del T} \right|_{T=T_n} \right) \fineq{,}
	\label{eq:pt:alpha}
\end{align}
where $\Delta V$ was defined below \cref{eq:pt:hubble}.
Second, the transition's inverse duration $\beta$ is obtained via
\begin{align}
	\label{eq:pt:beta}
	\beta & = H(T_n) T_n \cdot \left. \frac{\dd (S_3/T)}{\dd T} \right|_{T=T_n} \fineq{.} \\
	\intertext{In order to make sure that the use of the above definition of $\beta$ is justified, we follow Ref.~\cite{Megevand2017} and additionally check whether the quantity}
	\label{eq:pt:betaPrime}
	{\beta^\prime}^2 & := \frac{1}{2} H(T_n)^2 T_n^2 \cdot \left. \frac{\dd^2 (S_3/T)}{\dd T^2} \right|_{T=T_n}
\end{align}
is always small compared to $\beta$.
Indeed, we find that \mbox{$\beta \gg \beta^\prime$} (as well as \mbox{$\beta/H \gg 1$}) for all investigated points.
Just as the nucleation temperature, the quantities $\alpha$ and $\beta$ are vital to determine the gravitational wave signal and are listed in \cref{tab:paraGW} for our benchmark points of \cref{tab:bench}.

\begin{figure}[t]
	\centering
	\includegraphics[scale=0.92]{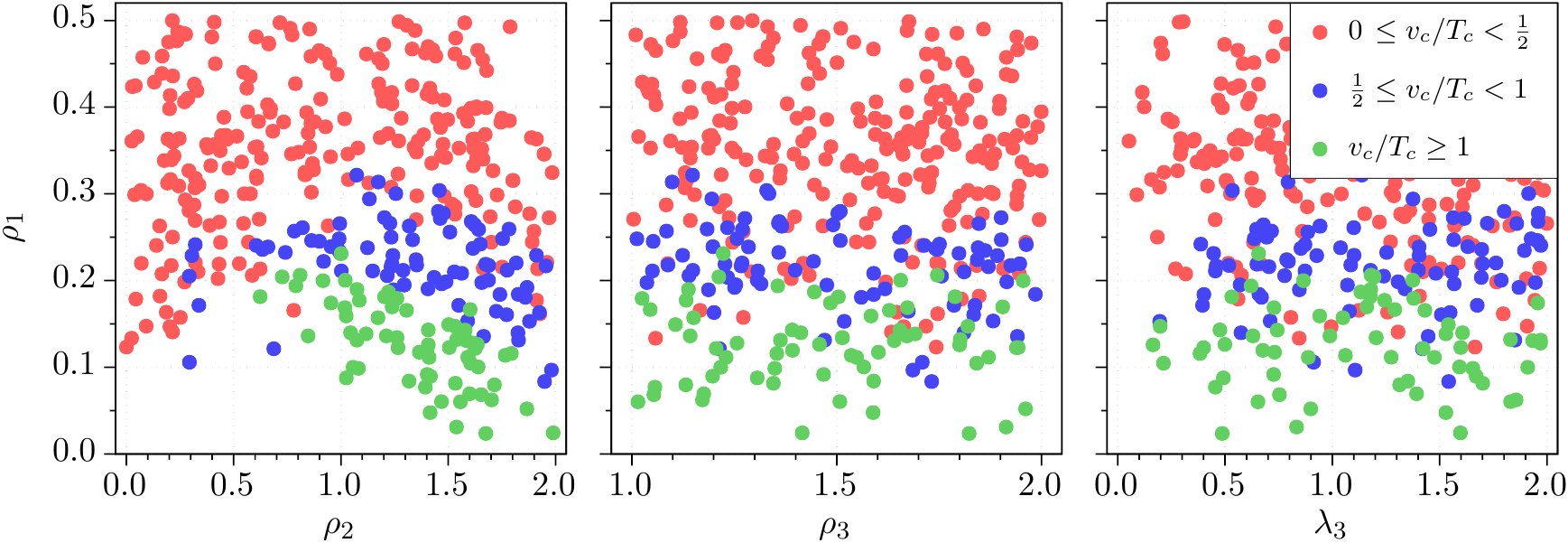}
	\caption{Strength of the LR-breaking phase transition as measured by the ratio $v_c/T_c$ with \mbox{$v_c:=v_R(T_c)$}. The randomly generated parameter points are displayed in two-dimensional projections of the full parameter space defined in \cref{eq:para_space}. The strongest transitions correspond to the green data points.}
	\label{fig:pt:correlations}
\end{figure}

As exemplified by the results for our benchmarks BP1 and BP2, a generic%
\footnote{In the sense that none of the non-zero, dimensionless couplings is particularly small, \cf \cref{eq:para_space}.}
parameter point of the considered left-right symmetric model predicts a relatively fast (large $\beta$), but weak (small $\alpha$) LR-breaking phase transition.
The gravitational wave spectrum associated with such a transition turns out to be out of reach of any current or proposed observatory, see \cref{sec:gw}.
Still, it is worthwhile to investigate whether there is some part of the multi-dimensional parameter space defined in \cref{eq:para_space} which is more promising in terms of PT strength \textendash\ and thus GW signal amplitude \textendash\ than others.
Since calculating $\alpha$ for a large number of benchmark points is computationally expensive, we employ an alternative measure to quantify the transition's strength for this purpose, namely the non-trivial VEV at the critical temperature normalized by the latter, $v_c/T_c$, where we abbreviated \mbox{$v_c:=v_R(T_c)$}.
In \cref{fig:pt:correlations} we show $v_c/T_c$ for our benchmark points in particularly interesting two-dimensional projections of the full parameter space.
As is clearly visible strong phase transitions with \mbox{$v_c/T_c\geq 1$} only occur for sufficiently small $\rho_1$ irrespective of the other couplings' values.
Investigating all possible parameter combinations in an analogous manner, we did not find any similarly significant correlation between the PT strength and a (dimensionless) coupling other than $\rho_1$.
In total, this motivates our choice of \mbox{$\rho_1=\mathcal{O}(\num{e-3})$} for our benchmarks BP3 and BP4, \cf \cref{tab:bench}.

From a physics perspective the importance of small $\rho_1$ to facilitate strong LR-breaking phase transitions can be understood by starting with the observation that it is $\rho_1$ which crucially shapes the model's tree-level potential in the $\dr$ field direction, \cf \cref{eq:V0simple}.
On the one hand, it governs the size of the quartic term.
On the other hand, and more importantly, it also sets the size of the triplet mass term $\mu_3^2$.
More precisely, for a fixed value of $v_R$, the tadpole equation \eqref{eq:x-9} entails the following implication
\begin{align}
	\rho_1 \ll 1
	\qquad\Longrightarrow\qquad
	\mu_3 \ll v_R \fineq{.}
\end{align}
In other words, choosing $\rho_1$ to be small brings the model's $\dr$ sector near its classically scale-invariant limit \mbox{$\mu_3/v_R \to 0$}.
Now, it is well-known that phase transitions in theories based on nearly conformal dynamics are typically strong and of first order \cite{Konstandin:2011dr,Megevand2017}, which explains the importance of the $\rho_1$ coupling for the strength of the gravitational wave signal.

\section{Gravitational wave signature}
\label{sec:gw}
\noindent
Following the detailed description of the model and the analysis of the phase transition, we are now fully equipped to discuss the main results of this work and finally present the gravitational wave signature associated with the left-right symmetry breaking first-order phase transition. The prerequisite for gravitational wave production is spherical symmetry breaking of the bubbles that contain a true vacuum. This is what occurs in their collisions and there are generally three distinct sources for gravitational wave production: collisions of bubble shells \cite{Kosowski}, sound waves \cite{Hindmarsh:2013xza} and magnetohydrodynamic turbulence \cite{Caprini:2006jb} in the plasma. For all of our benchmark points we compared  $\alpha$ with $\alpha_\infty$ parameter (defined in Eq.~(25) of \cite{Caprini:2015zlo}) and found that the latter is always larger by at least a factor of few. This implies that nucleated bubbles expand in what is usually referred to as a ``non-runaway" scenario \cite{Caprini:2015zlo}. In such case only the contributions from the sound waves and magnetohydrodynamic turbulence in the plasma can induce relevant gravitation wave signature. Hence, the total gravitational wave strength can be written as the sum of these two components
\begin{align}
\Omega_\text{\tiny{GW}} h^2 \simeq \Omega_\text{sw}\, h^2 + \Omega_\text{turb}\, h^2\,,
\label{eq:GWcont}
\end{align}
where
\begin{align}
\Omega_\text{sw} h^2 = 2.65\cdot 10^{-6} \left(\frac{H}{\beta}\right)\left(\frac{\kappa_v \,\alpha}{1+\alpha} \right)^2 \left(\frac{100}{g_*}\right)^{1/3} v_w \left(\frac{f}{f_\text{sw}}\right)^3 \left(\frac{7}{4+3\,(f/f_\text{sw})^2}\right)^{7/2}\,,
\label{eq:SW}
\end{align}
is the contribution arising from sound waves and
\begin{align}
\Omega_\text{turb} h^2 = 3.35\cdot 10^{-4}\, \left(\frac{H}{\beta}\right) \left(\frac{\kappa_\text{turb} \,\alpha}{1+\alpha} \right)^{3/2} \left(\frac{100}{g_*}\right)^{1/3} v_w \left(\frac{f}{f_\text{turb}}\right)^3 \frac{1}{\big[1+(f/f_\text{turb})\big]^{11/3}\,(1+8\pi f/h_*)}\,,
\label{eq:TURB}
\end{align}
stems from magnetohydrodynamic turbulence. 
In \cref{eq:SW}, $\kappa_v$ represents the efficiency for the conversion of latent heat into the bulk motion which yields \cite{Caprini:2015zlo} 
\begin{align}
\kappa_v =  \alpha \left(0.73+0.083 \sqrt{\alpha} + \alpha \right)^{-1}\,,
\label{eq:kv}
\end{align}
for the assumed bubble wall velocity of $v_w=1$. The peak frequency of the sound wave contribution, $f_\text{sw}$, equals
\begin{align}
f_\text{sw}=1.9\cdot 10^{-5}\, v_w^{-1} \left(\frac{\beta}{H}\right) \left(\frac{T_n}{100\,\text{GeV}}\right) \left(\frac{g_*}{100}\right)^{1/6} \,\text{Hz}\,.
\label{eq:fSW}
\end{align}
In \cref{eq:TURB},
\begin{align}
h_*=16.5\cdot 10^{-6} \left(\frac{T_n}{100\,\text{GeV}}\right)  \left(\frac{g_*}{100}\right)^{1/6} \,\text{Hz}\,,
\label{eq:h_star}
\end{align}
$\kappa_\text{turb}=0.05 \, \kappa_v$ \cite{Caprini:2015zlo} and the peak frequency is
\begin{align}
f_\text{turb}=2.7\cdot 10^{-5} \,v_w^{-1} \left(\frac{\beta}{H}\right) \left(\frac{T_n}{100\,\text{GeV}}\right) \left(\frac{g_*}{100}\right)^{1/6} \,\text{Hz}\,.
\label{eq:fTURB}
\end{align}
 Note that the previously introduced parameters $\alpha$, $\beta$ and $T_n$ also enter in the expression for the gravitational wave spectrum (in addition, $\alpha$ also contributes indirectly through $\kappa_v$ and $\kappa_\text{turb}$). The values of these parameters for the four considered benchmark points are given in \cref{tab:paraGW}. 

\def\arraystretch{1.6}
\begin{table}[b]
	\sisetup{round-mode=figures}
	\begin{tabular}{l@{\hskip 0.3cm}S[table-format=1.4,round-precision=2]S[table-format=4.1,round-precision=4]S[table-format=4.1,round-precision=4]S[table-format=4.1,round-precision=4]}
	\toprule
		   & {\symh{2em}{$\alpha$}} & {\symh{1em}{$\beta/H$}} & {\symh{1em}{$T_n$ [GeV]}} & {\symh{1em}{$T_c$ [GeV]}} \\
	\cline{2-5}
	BP1  & 0.0035   & 4006.84 & 5896.12 &  6216.1  \\
	BP2  & 0.0034   & 3457.95 & {\num{5.7536087e5}} & {\num{6.0631e5}} \\
	BP3  & 0.4574   & 626.17 & 608.28  & 9451.1 \\
	BP4  & 0.1844      &  1385.80 & 4484.2  &  7469.2 \\
	\botrule
	\end{tabular}
	\caption{Values of $\alpha$, $\beta/H$, $T_n$ and $T_c$ for the benchmark points considered in this work.}
	\label{tab:paraGW}
\end{table}

\begin{figure}[t]
	\centering
	\includegraphics[scale=0.53]{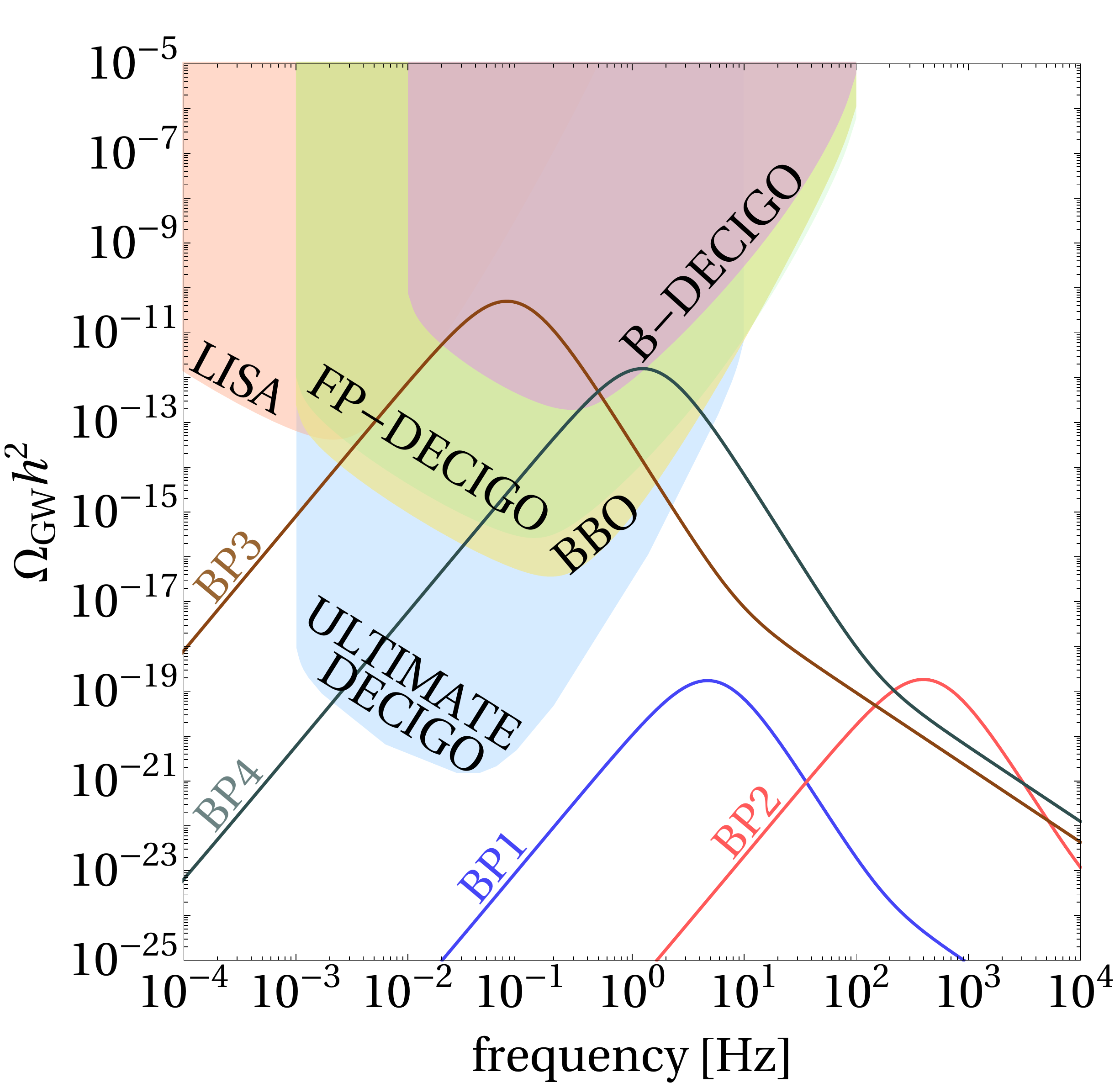}
	\caption{Stochastic gravitational wave spectra for the benchmark points given in \cref{tab:bench}. We also show the sensitivity curves of LISA \cite{Caprini:2015zlo}, BBO \cite{Corbin:2005ny} and three stages of the DECIGO experiment \cite{Seto:2001qf,Kuroyanagi:2014qaa}. The power-law integrated sensitivity curves \cite{Thrane:2013oya} are calculated assuming a runtime of $t_\text{obs}=5$ years and threshold SNR equal to 10 for a number of forthcoming space-based detectors. As is evident from the figure, the gravitational wave spectra for BP3 and BP4 intersect several sensitivity curves which indicates that such experiments would either lead to the discovery or successfully exclude these benchmark points.}
	\label{fig:spectrum}
\end{figure}

In \cref{fig:spectrum} we show the gravitational wave spectra for all of our benchmark points and confront them with the sensitivities of several future space-based detectors.
According to the shown spectra,  BP1 and BP2 (benchmark points arising from the general scan without particularly small $\rho_1$) are beyond the reach of all considered experiments.
For these two points, the strength of produced gravitational waves surpasses the maximal sensitivity reach of ULTIMATE DECIGO; however, due to large $\beta$ (see \cref{tab:paraGW}), the peak frequency exhibits a shift to the region in which none of the considered detectors operate.
Following the identification of the correlation between parameters in the scalar potential with respect to the strength of phase transition (see again \cref{fig:pt:correlations}), our more refined parameter scan identified BP3 and BP4 for which, as obvious from \cref{fig:spectrum}, the testability is guaranteed. For BP3 and BP4, the parameters of the scalar potential yield a tree-level value of $v_R=10$ TeV and $v_R=50$ TeV, respectively.
As discussed in the previous sections, we tune the right-handed neutrino Yukawa coupling $y_M$ in order to attain the same value at the radiative level.
Generally, it is not a problem if radiative effects significantly contribute to the potential and alter the tree-level vacuum structure.
However, in order to simplify our numerical treatment, we have imposed dominance of the tree-level potential (\cf also the discussion at the end of \cref{sec:benchmarks}).
Let us stress that we explicitly checked that similarly strong GW signals can likewise arise in situations where still \mbox{$\rho_1\lesssim\mathcal{O}(\num{e-3})$}, but where $y_M$ is no longer tuned to maintain a given tree-level vacuum also at loop level.
The left-right breaking scale is then predominantly set by the Coleman-Weinberg potential so that the reliability of benchmark points obtained from tree-level calculations must be called into question.
Still, it serves to demonstrate that the occurrence of a strong gravitational wave signal depends on the smallness of $\rho_1$ rather than on a particularly tuned value of $y_M$.

From \cref{fig:spectrum} one can also infer, for BP3 and BP4, the change of the frequency dependence of the spectrum at around $10-100$ Hz. This is the region in which the sound wave contribution, that is dominant at smaller frequencies, becomes suppressed with respect to the component stemming from magnetohydrodynamical turbulence in the plasma. This can also be inferred analytically from \cref{eq:SW,eq:TURB} from where it is obvious that the sound wave contribution decreases more strongly at larger frequencies. Namely, for the contribution from sound waves (magnetohydrodynamic turbulence) we have $\Omega_\text{sw} h^2 \propto f^{-4}$ ($\Omega_\text{turb} h^2 \propto f^{-5/3}$) in the limit $f\to \infty$.

In order to provide more quantitative information on the potential for discovering gravitational waves in the considered left-right symmetric model we define the signal-to-noise ratio (SNR) \cite{Thrane:2013oya}
\begin{align}
	\text{SNR}=\sqrt{2 t_{\text{obs}} \int_{f_{\text{min}}}^{f_{\text{max}}} \!\! \mbox{d} f \, \bigg[
\frac{\Omega_\text{\tiny{GW}}(f)\, h^2}{\Omega_\text{noise}(f)\, h^2}\bigg]^2}\,.
	\label{eq:SNR}
\end{align}
Here, $t_\text{obs}$ is the runtime of a given experiment in seconds, $\Omega_\text{noise} h^2$ is the effective strain noise power spectral density \cite{Moore:2014lga} and the integral goes in the range of frequencies $(f_\text{min}, f_\text{max})$ in which a given experiment is sensitive. Note also that for single-detector configurations (LISA and B-DECIGO) the $\sqrt{2}$ from \cref{eq:SNR} should be dropped. Various experiments report different threshold values of SNR for which a detection is guaranteed (see \cite{Brdar:2018num} and references therein) and typically values between $5$ and $10$ are reported, the latter being a conservative estimate.
For BP1 and BP2 clearly SNR values for all experiments are lower than 1.
BP3 and BP4 offer a much more promising discovery potential: for both of these benchmark points, SNR exceeds $10^3$ for FP-DECIGO, BBO and ULTIMATE DECIGO experiments. In addition, B-DECIGO would also lead to a successful discovery since SNR for BP3 (BP4) is $672.70$ ($16.69$). At the moment, among all the proposed space-based detectors, LISA is most mature and is expected to launch in 2034.
LISA will not be sensitive to gravitational wave spectra associated to BP4, but, for BP3, SNR value of $11.76$ is obtained. This clearly indicates that this parameter point will be probed.  

In summary, we have demonstrated in the framework of the considered left-right model, that gravitational wave detectors will in the near future serve as a powerful complementary probe to the existing terrestrial searches. We have shown this explicitly by presenting benchmark points which are testable at forthcoming space-based detectors such as LISA, but beyond the reach of LHC.

\section{Summary and Conclusions}
\label{sec:summary}
\noindent
In the present work we studied the observational prospects for gravitational waves in the framework of the minimal left-right symmetric model with scalar triplets.
We found that the parity-breaking phase transition associated with $SU(2)_L\times SU(2)_R\times U(1)_{B-L} \to SU(2)_L\times U(1)_Y$ is of first order in a vast portion of parameter space and may therefore indeed result in the production of a stochastic gravitational wave background.
However, it turns out that the amplitude of the predicted spectrum is typically too small to be probed at any of the currently proposed space-based experiments.
Nevertheless, a thorough investigation of the model's parameter space revealed that there exists a systematic correlation between one of the couplings in the scalar potential (namely $\rho_1$) and the strength of the generated gravitational wave signal.
To be more precise, we found that the left-right transition can produce sufficiently strong and thus testable signals, if one accepts a moderate amount of parameter tuning reducing $\rho_1$ to be of order \num{e-3}.
From a \textit{physics} point of view, the above-described behavior can be understood by realizing that the limit \mbox{$\rho_1 \to 0$} entails the development of a flat direction in the model's tree-level potential along the right-handed scalar triplet's field axis.
The observed correlation then follows from well-established results on theories based on nearly conformal dynamics and classical scale invariance, namely that phase transitions in such models are generically strong and of first order.
On the downside, small values for $\rho_1$ and the resulting shallow tree-level potential lead to a \textit{technical} difficulty:
The scale $v_R$ of left-right symmetry breaking usually changes substantially when going from the tree- to the one-loop order thus calling into question the reliability of benchmark points obtained from tree-level calculations.
This issue can be circumvented by either determining benchmark points with full one-loop precision, or by explicitly requiring the Coleman-Weinberg contribution not to change the tree-level vacuum.
For the sake of simplicity, we chose the latter option, which was achieved by employing mild fine-tuning of the right-handed neutrino Yukawa coupling. However, we argued that the appearance of a strong gravitational wave signal depends on the smallness of $\rho_1$ rather than on a particularly tuned form of the Coleman-Weinberg potential.

In the near future, gravitational wave searches will thus provide a novel and complementary probe not only for the mechanism of left-right symmetry breaking, but also for the generation of neutrino masses, which is embedded in the considered model.
While searches at colliders have limitations on the energies that can be tested, data from gravitational wave observatories may lead to the discovery of left-right symmetry breaking at higher scales.

\section*{Acknowledgments}
\noindent
VB would like to thank Miha Nemev\v{s}ek for very useful discussions during the MITP workshop \enquote{Indirect Searches for New Physics Across the Scales}. In addition, we would like to thank Toby Opferkuch for sharing sensitivity curves of the ULTIMATE DECIGO phase.

\appendix
\section{Particle Mass Spectra}
\label{app:spectra}
\noindent
In this appendix we list the field-dependent tree-level mass-squared matrices and their corresponding eigenvalues for the particle content of the considered left-right symmetric model in the limit of large field values \mbox{$\dr=\operatorname{Re}\delta_R^0/\sqrt{2}$}.

\subsection{Scalars}
\noindent
In the assumed limit, the mass-squared matrix of the scalar bi-doublet $\Phi$ is block diagonal with the same \mbox{$2\times2$} block repeated four times.
Thus, it is fully characterized by the aforementioned \mbox{$2\times 2$} matrix, which reads
\begin{align}
	\mathcal{M}^2_\Phi(\dr) & =
	\begin{pmatrix}
		-\mu_1^2 + \tfrac{1}{2}\alpha_1 \dr^2 & -2\mu_2^2 + \alpha_2\dr^2 \\
		-2\mu_2^2 + \alpha_2\dr^2 & -\mu_1^2 + \tfrac{1}{2}(\alpha_1+\alpha_3) \dr^2 \\
	\end{pmatrix}
	\fineq{,}
	\label{eq:spectra:bi-doublet}
\end{align}
with the eigenvalues
\begin{align}
m^2_{\Phi}(\dr) & = \frac{1}{2}\alpha_1\dr^2\pm\frac{1}{4}\sqrt{16\alpha_2^2\dr^4-64\alpha_2\mu_2^2\dr^2+\alpha_3^2\dr^4+64\mu_2^4}+\alpha_3\dr^2-4\mu_1^2
\fineq{,}
\end{align}
each of which has a multiplicity of four.
Next, the left-handed triplet has a diagonal mass-squared matrix with six equal eigenvalues of the form
\begin{align}
	m^2_{\Delta_L}(\dr) & = -\mu_3^2 + \tfrac{1}{2}\rho_3 \dr^2 \fineq{.}
	\label{eq:spectra:DeltaL}
	\intertext{Lastly, in the case of the right-handed triplet, the mass-squared matrix is again diagonal; however, there are three different eigenvalues. The first one,}
	m^2(\dr) & = -\mu_3^2 + 3\rho_1 \dr^2 \fineq{,}
	\intertext{appears only once and it corresponds to the neutral CP-even component. Next, there is a three times degenerate eigenvalue}
	m^2(\dr) & = -\mu_3^2 + \rho_1 \dr^2 \fineq{,}
	\intertext{respective to the neutral CP-odd and singly-charged components. Finally, the last two eigenvalues of the form}
	m^2(\dr) & = -\mu_3^2 + (\rho_1 + 2\rho_2) \dr^2
\end{align}
give the masses of the doubly-charged fields.

\subsection{Gauge Fields}
\noindent
On the one hand, the charged right-handed gauge bosons have the following field-dependent masses
\begin{align}
	m^2_{W_R}(\dr) & = \tfrac{1}{4}g^2 \dr^2 \fineq{.}
\end{align}
On the other hand, the mass-squared matrix of neutral gauge bosons in the basis $(W_L^3, W_R^3, B)$ and again assuming large $\dr$ reads
\begin{align}
	\mathcal{M}^2_{\text{neutral}}(\dr) = \frac{\dr^2}{4}
	\begin{pmatrix}
		0 & 0 & 0 \\
		0 & g^2 & -g g_{B-L} \\
		0 & -g g_{B-L} & g_{B-L}^2 \\
	\end{pmatrix}
	\fineq{.}
	\label{eq:spectra:neutral}
\end{align}
Two eigenvalues of the above mass-squared matrix are zero, the remaining one is given by
\begin{align}
	m^2_{Z_R}(\dr) & = \tfrac{1}{4}(g^2+g_{B-L}^2)\dr^2 \fineq{.}
\end{align}

\subsection{Fermions}
\noindent
The three right-handed neutrinos acquire the field-dependent mass
\begin{align}
	m_{\nu_R}^2(\dr) = 2 y_M^2\dr^2 \fineq{.}
\end{align}
In the limit of large $\dr$ this is the only relevant fermionic mass that needs to be accounted for.

\section{Thermal Self-Energies}
\label{app:thermalmasses}
\noindent
Here, we list the bosonic thermal self-energies $\Pi_{i}$ needed to calculate the daisy improvement of the finite-temperature effective potential in \cref{eq:VD}.
They can be obtained as the leading temperature-dependent self-energy contributions, which are proportional to $T^2$ \cite{Carrington:1991hz}.
The resulting field-dependent thermal masses $M^2(r,T)$ are then obtained as the eigenvalues of the sum of the thermal self-energy matrices provided in the following and the tree-level mass-squared matrices as specified in Eqs.~\eqref{eq:spectra:bi-doublet} and \eqref{eq:spectra:DeltaL} to \eqref{eq:spectra:neutral} of \cref{app:spectra}.

\subsection{Scalar Bosons}
\noindent
The thermal self-energies corresponding to the bi-doublet and the triplet scalars, respectively, can be summarized in a compact way following the parametrization in \cref{eq:x-1} as
\begin{align}
	\Pi_\Phi(T) & = \frac{T^2}{24}
	\begin{pmatrix}
		20\lambda_1 + 8\lambda_3 + 12\alpha_1 + 6\alpha_3 + 9g^2 + 6y_t^2 & 24(\lambda_4 + \alpha_2) \\
		24(\lambda_4 + \alpha_2) & 20\lambda_1 + 8\lambda_3 + 12\alpha_1 + 6\alpha_3 + 9g^2 + 6y_t^2  \\
	\end{pmatrix}
	\fineq{,} \\
	\Pi_\Delta(T) & = \frac{T^2}{12} (8\rho_1 + 4\rho_2 + 3\rho_3 + 4\alpha_1 + 2\alpha_3 + 6g^2 + 3g^2_\mathsmaller{B-L} + 6y_M^2) \fineq{.}
\end{align}
The above expressions incorporate contributions from scalars' self-interactions as well as contributions induced by gauge fields and fermions. As argued in Sec.~\ref{sec:model}, for our analysis it is enough to take into account only the third fermionic generation; hence, we include here only the contributions from the top quark and the right-handed neutrino. As in the SM, the terms in \cref{eq:yukDoublet} contribute to the thermal self-energies of all (real) bi-doublet components as $\Pi_\Phi\supseteq y_t^2T^2/4$. The Yukawa terms for the heavy right-handed neutrino in \cref{eq:yukTriplet} then give the contribution to the thermal self-energies of all (real) triplet components as $\Pi_\Delta\supseteq y_M^2T^2/2$.

\subsection{Gauge Bosons}
\noindent
As for the gauge fields, only their longitudinal components contribute to the daisy improvement in \cref{eq:VD}.
The thermal self-energy for the charged right-handed gauge bosons is
\begin{align}
\Pi^L_{W_R}=\tfrac{3}{2} g^2 T^2 \fineq{,}
\end{align}
while the thermal self-energy matrix for the neutral gauge fields in the basis $(W_L^3, W_R^3, B)$ is
\begin{align}
\Pi^L_{\text{neutral}}=  \frac{T^2}{6}  \,\operatorname{diag}\left(9 \,g^2,\, 9\, g^2,\, 17 \,g_{B-L}^2 \right) \fineq{.}
\end{align}

\bibliographystyle{JHEP}
\bibliography{refs}

\providecommand{\href}[2]{#2}\begingroup\raggedright\begin{thebibliography}{10}

\bibitem{Pati/Salam}
J.~C. Pati and A.~Salam, {\it Lepton number as the fourth \enquote{color}},
  {\em Phys. Rev. D} {\bf 10} (1974) 275.

\bibitem{Goran/Mohapatra}
G.~Senjanovic and R.~N. Mohapatra, {\it Exact left-right symmetry and
  spontaneous violation of parity},  {\em Phys. Rev. D} {\bf 12} (1975) 1502.

\bibitem{Mohapatra/Pati}
R.~N. Mohapatra and J.~C. Pati, {\it Left-right gauge symmetry and an
  \enquote{isoconjugate} model of cp violation},  {\em Phys. Rev. D} {\bf 11}
  (1975) 566.

\bibitem{Mohapatra/Pati2}
R.~N. Mohapatra and J.~C. Pati, {\it \enquote{Natural} left-right symmetry},
  {\em Phys. Rev. D} {\bf 11} (1975) 2558.

\bibitem{MS}
R.~N. Mohapatra and D.~P. Sidhu, {\it Gauge theories of weak interactions with
  left-right symmetry and the structure of neutral currents},  {\em Phys. Rev.
  D} {\bf 16} (1977) 2843.

\bibitem{Maiezza:2016ybz}
A.~Maiezza, G.~Senjanovic, and J.~C. Vasquez, {\it {Higgs sector of the minimal
  left-right symmetric theory}},  {\em Phys. Rev.} {\bf D95} (2017), no.~9
  095004, [\href{http://www.arxiv.org/abs/1612.09146}{{\tt 1612.09146}}].

\bibitem{Deshpande:1990ip}
N.~G. Deshpande, J.~F. Gunion, B.~Kayser, and F.~I. Olness, {\it {Left-right
  symmetric electroweak models with triplet Higgs}},  {\em Phys. Rev.} {\bf
  D44} (1991) 837--858.

\bibitem{Senjanovic:2016bya}
G.~Senjanovic, {\it {Is left–right symmetry the key?}},  {\em Mod. Phys.
  Lett.} {\bf A32} (2017), no.~04 1730004,
  [\href{http://www.arxiv.org/abs/1610.04209}{{\tt 1610.04209}}].

\bibitem{Brdar:2018sbk}
V.~Brdar and A.~Y. Smirnov, {\it {Low Scale Left-Right Symmetry and Naturally
  Small Neutrino Mass}},  {\em JHEP} {\bf 02} (2019) 045,
  [\href{http://www.arxiv.org/abs/1809.09115}{{\tt 1809.09115}}].

\bibitem{FileviezPerez:2016erl}
P.~Fileviez~Perez, C.~Murgui, and S.~Ohmer, {\it {Simple Left-Right Theory:
  Lepton Number Violation at the LHC}},  {\em Phys. Rev.} {\bf D94} (2016),
  no.~5 051701, [\href{http://www.arxiv.org/abs/1607.00246}{{\tt 1607.00246}}].

\bibitem{Goran}
R.~N. Mohapatra and G.~Senjanovi\ifmmode~\acute{c}\else \'{c}\fi{}, {\it
  Neutrino mass and spontaneous parity nonconservation},  {\em Phys. Rev.
  Lett.} {\bf 44} (1980) 912.

\bibitem{Yanagida:1979as}
T.~Yanagida, {\it {Horizontal Symmetry and Masses of Neutrinos}},  {\em Conf.
  Proc.} {\bf C7902131} (1979) 95.

\bibitem{GellMann:1980vs}
M.~Gell-Mann, P.~Ramond, and R.~Slansky, {\it {Complex Spinors and Unified
  Theories}},  {\em Conf. Proc.} {\bf C790927} (1979) 315,
  [\href{http://www.arxiv.org/abs/1306.4669}{{\tt 1306.4669}}].

\bibitem{Minkowski}
P.~Minkowski, {\it $\mu\to e\gamma$ at a rate of one out of $10^9$ muon
  decays?},  {\em Physics Letters B} {\bf 67} (1977), no.~4 421.

\bibitem{Fritzsch:1974nn}
H.~Fritzsch and P.~Minkowski, {\it {Unified Interactions of Leptons and
  Hadrons}},  {\em Annals Phys.} {\bf 93} (1975) 193--266.

\bibitem{Georgi:1974my}
H.~Georgi, {\it {The State of the Art—Gauge Theories}},  {\em AIP Conf.
  Proc.} {\bf 23} (1975) 575--582.

\bibitem{Arbelaez:2013nga}
C.~Arbeláez, M.~Hirsch, M.~Malinský, and J.~C. Romão, {\it {LHC-scale
  left-right symmetry and unification}},  {\em Phys. Rev.} {\bf D89} (2014),
  no.~3 035002, [\href{http://www.arxiv.org/abs/1311.3228}{{\tt 1311.3228}}].

\bibitem{Deppisch:2017xhv}
F.~F. Deppisch, T.~E. Gonzalo, and L.~Graf, {\it {Surveying the SO(10) Model
  Landscape: The Left-Right Symmetric Case}},  {\em Phys. Rev.} {\bf D96}
  (2017), no.~5 055003, [\href{http://www.arxiv.org/abs/1705.05416}{{\tt
  1705.05416}}].

\bibitem{Chen:2013fna}
C.-Y. Chen, P.~S.~B. Dev, and R.~N. Mohapatra, {\it {Probing Heavy-Light
  Neutrino Mixing in Left-Right Seesaw Models at the LHC}},  {\em Phys. Rev.}
  {\bf D88} (2013) 033014, [\href{http://www.arxiv.org/abs/1306.2342}{{\tt
  1306.2342}}].

\bibitem{Lindner:2016lxq}
M.~Lindner, F.~S. Queiroz, W.~Rodejohann, and C.~E. Yaguna, {\it {Left-Right
  Symmetry and Lepton Number Violation at the Large Hadron Electron Collider}},
   {\em JHEP} {\bf 06} (2016) 140,
  [\href{http://www.arxiv.org/abs/1604.08596}{{\tt 1604.08596}}].

\bibitem{Patra:2015bga}
S.~Patra, F.~S. Queiroz, and W.~Rodejohann, {\it {Stringent Dilepton Bounds on
  Left-Right Models using LHC data}},  {\em Phys. Lett.} {\bf B752} (2016)
  186--190, [\href{http://www.arxiv.org/abs/1506.03456}{{\tt 1506.03456}}].

\bibitem{Dev:2015kca}
P.~S.~B. Dev, D.~Kim, and R.~N. Mohapatra, {\it {Disambiguating Seesaw Models
  using Invariant Mass Variables at Hadron Colliders}},  {\em JHEP} {\bf 01}
  (2016) 118, [\href{http://www.arxiv.org/abs/1510.04328}{{\tt 1510.04328}}].

\bibitem{Rodriguez:2002ey}
Y.~Rodriguez and C.~Quimbay, {\it {Spontaneous CP phases and flavor changing
  neutral currents in the left-right symmetric model}},  {\em Nucl. Phys.} {\bf
  B637} (2002) 219--242, [\href{http://www.arxiv.org/abs/hep-ph/0203178}{{\tt
  hep-ph/0203178}}].

\bibitem{Nemevsek:2018bbt}
M.~Nemev\v{s}ek, F.~Nesti, and G.~Popara, {\it {Keung-Senjanovi\'c process at
  the LHC: From lepton number violation to displaced vertices to invisible
  decays}},  {\em Phys. Rev.} {\bf D97} (2018), no.~11 115018,
  [\href{http://www.arxiv.org/abs/1801.05813}{{\tt 1801.05813}}].

\bibitem{Witten1984}
E.~Witten, {\it {Cosmic Separation of Phases}},  {\em Phys. Rev.} {\bf D30}
  (1984) 272--285.

\bibitem{Hogan1984}
C.~J. Hogan, {\it {Nucleation of cosmological phase transitions}},  {\em Phys.
  Lett.} {\bf 133B} (1983) 172--176.

\bibitem{Hogan1986b}
C.~J. Hogan, {\it {Gravitational radiation from cosmological phase
  transitions}},  {\em Mon. Not. Roy. Astron. Soc.} {\bf 218} (1986) 629--636.

\bibitem{Turner1990a}
M.~S. Turner and F.~Wilczek, {\it {Relic gravitational waves and extended
  inflation}},  {\em Phys. Rev. Lett.} {\bf 65} (1990) 3080--3083.

\bibitem{Kamionkowski1993}
M.~Kamionkowski, A.~Kosowsky, and M.~S. Turner, {\it {Gravitational radiation
  from first order phase transitions}},  {\em Phys. Rev.} {\bf D49} (1994)
  2837--2851, [\href{http://www.arxiv.org/abs/astro-ph/9310044}{{\tt
  astro-ph/9310044}}].

\bibitem{Grojean2007}
C.~Grojean and G.~Servant, {\it {Gravitational Waves from Phase Transitions at
  the Electroweak Scale and Beyond}},  {\em Phys. Rev.} {\bf D75} (2007)
  043507, [\href{http://www.arxiv.org/abs/hep-ph/0607107}{{\tt
  hep-ph/0607107}}].

\bibitem{Ellis:2018mja}
J.~Ellis, M.~Lewicki, and J.~M. No, {\it {On the Maximal Strength of a
  First-Order Electroweak Phase Transition and its Gravitational Wave Signal}},
   \href{http://www.arxiv.org/abs/1809.08242}{{\tt 1809.08242}}.
  [JCAP1904,003(2019)].

\bibitem{Kajantie1996a}
K.~Kajantie, M.~Laine, K.~Rummukainen, and M.~E. Shaposhnikov, {\it {The
  Electroweak phase transition: A Nonperturbative analysis}},  {\em Nucl.
  Phys.} {\bf B466} (1996) 189--258,
  [\href{http://www.arxiv.org/abs/hep-lat/9510020}{{\tt hep-lat/9510020}}].

\bibitem{Kajantie1996b}
K.~Kajantie, M.~Laine, K.~Rummukainen, and M.~E. Shaposhnikov, {\it {Is there a
  hot electroweak phase transition at m(H) larger or equal to m(W)?}},  {\em
  Phys. Rev. Lett.} {\bf 77} (1996) 2887--2890,
  [\href{http://www.arxiv.org/abs/hep-ph/9605288}{{\tt hep-ph/9605288}}].

\bibitem{Aoki2006a}
Y.~Aoki, Z.~Fodor, S.~D. Katz, and K.~K. Szabo, {\it {The QCD transition
  temperature: Results with physical masses in the continuum limit}},  {\em
  Phys. Lett.} {\bf B643} (2006) 46--54,
  [\href{http://www.arxiv.org/abs/hep-lat/0609068}{{\tt hep-lat/0609068}}].

\bibitem{Aoki2006b}
Y.~Aoki, G.~Endrodi, Z.~Fodor, S.~D. Katz, and K.~K. Szabo, {\it {The Order of
  the quantum chromodynamics transition predicted by the standard model of
  particle physics}},  {\em Nature} {\bf 443} (2006) 675--678,
  [\href{http://www.arxiv.org/abs/hep-lat/0611014}{{\tt hep-lat/0611014}}].

\bibitem{Bhattacharya2014}
T.~Bhattacharya {\em et~al.}, {\it {QCD Phase Transition with Chiral Quarks and
  Physical Quark Masses}},  {\em Phys. Rev. Lett.} {\bf 113} (2014), no.~8
  082001, [\href{http://www.arxiv.org/abs/1402.5175}{{\tt 1402.5175}}].

\bibitem{Okada2018a}
N.~Okada and O.~Seto, {\it {Probing the seesaw scale with gravitational
  waves}},  {\em Phys. Rev.} {\bf D98} (2018), no.~6 063532,
  [\href{http://www.arxiv.org/abs/1807.00336}{{\tt 1807.00336}}].

\bibitem{Brdar:2018num}
V.~Brdar, A.~J. Helmboldt, and J.~Kubo, {\it {Gravitational Waves from
  First-Order Phase Transitions: LIGO as a Window to Unexplored Seesaw
  Scales}},  {\em JCAP} {\bf 1902} (2019) 021,
  [\href{http://www.arxiv.org/abs/1810.12306}{{\tt 1810.12306}}].

\bibitem{Dror:2019syi}
J.~A. Dror, T.~Hiramatsu, K.~Kohri, H.~Murayama, and G.~White, {\it {Testing
  Seesaw and Leptogenesis with Gravitational Waves}},
  \href{http://www.arxiv.org/abs/1908.03227}{{\tt 1908.03227}}.

\bibitem{Hasegawa:2019amx}
T.~Hasegawa, N.~Okada, and O.~Seto, {\it {Gravitational waves from the minimal
  gauged $U(1)_{B-L}$ model}},  {\em Phys. Rev.} {\bf D99} (2019), no.~9
  095039, [\href{http://www.arxiv.org/abs/1904.03020}{{\tt 1904.03020}}].

\bibitem{Schwaller2015}
P.~Schwaller, {\it {Gravitational Waves from a Dark Phase Transition}},  {\em
  Phys. Rev. Lett.} {\bf 115} (2015), no.~18 181101,
  [\href{http://www.arxiv.org/abs/1504.07263}{{\tt 1504.07263}}].

\bibitem{Jaeckel2016a}
J.~Jaeckel, V.~V. Khoze, and M.~Spannowsky, {\it {Hearing the signal of dark
  sectors with gravitational wave detectors}},  {\em Phys. Rev.} {\bf D94}
  (2016), no.~10 103519, [\href{http://www.arxiv.org/abs/1602.03901}{{\tt
  1602.03901}}].

\bibitem{Dev:2016feu}
P.~S.~B. Dev and A.~Mazumdar, {\it {Probing the Scale of New Physics by
  Advanced LIGO/VIRGO}},  {\em Phys. Rev.} {\bf D93} (2016), no.~10 104001,
  [\href{http://www.arxiv.org/abs/1602.04203}{{\tt 1602.04203}}].

\bibitem{Baldes2017}
I.~Baldes, {\it {Gravitational waves from the asymmetric-dark-matter generating
  phase transition}},  {\em JCAP} {\bf 1705} (2017), no.~05 028,
  [\href{http://www.arxiv.org/abs/1702.02117}{{\tt 1702.02117}}].

\bibitem{Baldes:2018emh}
I.~Baldes and C.~Garcia-Cely, {\it {Strong gravitational radiation from a
  simple dark matter model}},  {\em JHEP} {\bf 05} (2019) 190,
  [\href{http://www.arxiv.org/abs/1809.01198}{{\tt 1809.01198}}].

\bibitem{Breitbach:2018ddu}
M.~Breitbach, J.~Kopp, E.~Madge, T.~Opferkuch, and P.~Schwaller, {\it {Dark,
  Cold, and Noisy: Constraining Secluded Hidden Sectors with Gravitational
  Waves}},  {\em JCAP} {\bf 1907} (2019), no.~07 007,
  [\href{http://www.arxiv.org/abs/1811.11175}{{\tt 1811.11175}}].

\bibitem{Fairbairn:2019xog}
M.~Fairbairn, E.~Hardy, and A.~Wickens, {\it {Hearing without seeing:
  gravitational waves from hot and cold hidden sectors}},  {\em JHEP} {\bf 07}
  (2019) 044, [\href{http://www.arxiv.org/abs/1901.11038}{{\tt 1901.11038}}].

\bibitem{Helmboldt:2019pan}
A.~J. Helmboldt, J.~Kubo, and S.~van~der Woude, {\it {Observational prospects
  for gravitational waves from hidden or dark chiral phase transitions}},
  \href{http://www.arxiv.org/abs/1904.07891}{{\tt 1904.07891}}.

\bibitem{Croon:2018kqn}
D.~Croon, T.~E. Gonzalo, and G.~White, {\it {Gravitational Waves from a
  Pati-Salam Phase Transition}},  {\em JHEP} {\bf 02} (2019) 083,
  [\href{http://www.arxiv.org/abs/1812.02747}{{\tt 1812.02747}}].

\bibitem{Mohamadnejad:2019vzg}
A.~Mohamadnejad, {\it {Gravitational waves from scale-invariant vector dark
  matter model: Probing below the neutrino-floor}},
  \href{http://www.arxiv.org/abs/1907.08899}{{\tt 1907.08899}}.

\bibitem{Dev:2019njv}
P.~S.~B. Dev, F.~Ferrer, Y.~Zhang, and Y.~Zhang, {\it {Gravitational Waves from
  First-Order Phase Transition in a Simple Axion-Like Particle Model}},  {\em
  JCAP} {\bf 1911} (2019), no.~11 006,
  [\href{http://www.arxiv.org/abs/1905.00891}{{\tt 1905.00891}}].

\bibitem{Barenboim:1998ib}
G.~Barenboim and N.~Rius, {\it {Electroweak phase transitions in left-right
  symmetric models}},  {\em Phys. Rev.} {\bf D58} (1998) 065010,
  [\href{http://www.arxiv.org/abs/hep-ph/9803215}{{\tt hep-ph/9803215}}].

\bibitem{Choi:1992wb}
J.~Choi and R.~R. Volkas, {\it {The Effective potential at finite temperature
  in the left-right symmetric model}},  {\em Phys. Rev.} {\bf D48} (1993)
  1258--1265, [\href{http://www.arxiv.org/abs/hep-ph/9210223}{{\tt
  hep-ph/9210223}}].

\bibitem{Sagunski:2012pzo}
L.~Sagunski, {\em {Gravitational waves as cosmological probes for new physics
  between the electroweak and the grand-unification scale}}.
\newblock PhD thesis, Hamburg U., 2012.

\bibitem{Caprini:2015zlo}
C.~Caprini {\em et~al.}, {\it {Science with the space-based interferometer
  eLISA. II: Gravitational waves from cosmological phase transitions}},  {\em
  JCAP} {\bf 1604} (2016), no.~04 001,
  [\href{http://www.arxiv.org/abs/1512.06239}{{\tt 1512.06239}}].

\bibitem{Seto:2001qf}
N.~Seto, S.~Kawamura, and T.~Nakamura, {\it {Possibility of direct measurement
  of the acceleration of the universe using 0.1-Hz band laser interferometer
  gravitational wave antenna in space}},  {\em Phys. Rev. Lett.} {\bf 87}
  (2001) 221103, [\href{http://www.arxiv.org/abs/astro-ph/0108011}{{\tt
  astro-ph/0108011}}].

\bibitem{Corbin:2005ny}
V.~Corbin and N.~J. Cornish, {\it {Detecting the cosmic gravitational wave
  background with the big bang observer}},  {\em Class. Quant. Grav.} {\bf 23}
  (2006) 2435--2446, [\href{http://www.arxiv.org/abs/gr-qc/0512039}{{\tt
  gr-qc/0512039}}].

\bibitem{Dev:2018foq}
P.~S. Bhupal~Dev, R.~N. Mohapatra, W.~Rodejohann, and X.-J. Xu, {\it {Vacuum
  structure of the left-right symmetric model}},  {\em JHEP} {\bf 02} (2019)
  154, [\href{http://www.arxiv.org/abs/1811.06869}{{\tt 1811.06869}}].

\bibitem{Chauhan:2019fji}
G.~Chauhan, {\it {Vacuum Stability and Symmetry Breaking in Left-Right
  Symmetric Model}},  \href{http://www.arxiv.org/abs/1907.07153}{{\tt
  1907.07153}}.

\bibitem{Nemevsek:2016enw}
M.~Nemev\v{s}ek, F.~Nesti, and J.~C. Vasquez, {\it {Majorana Higgses at
  colliders}},  {\em JHEP} {\bf 04} (2017) 114,
  [\href{http://www.arxiv.org/abs/1612.06840}{{\tt 1612.06840}}].

\bibitem{Linde1981}
A.~D. Linde, {\it {Decay of the False Vacuum at Finite Temperature}},  {\em
  Nucl. Phys.} {\bf B216} (1983) 421. [Erratum: Nucl. Phys.B223,544(1983)].

\bibitem{Wainwright2012}
C.~L. Wainwright, {\it {CosmoTransitions: Computing Cosmological Phase
  Transition Temperatures and Bubble Profiles with Multiple Fields}},  {\em
  Comput. Phys. Commun.} {\bf 183} (2012) 2006--2013,
  [\href{http://www.arxiv.org/abs/1109.4189}{{\tt 1109.4189}}].

\bibitem{Espinosa:2010hh}
J.~R. Espinosa, T.~Konstandin, J.~M. No, and G.~Servant, {\it {Energy Budget of
  Cosmological First-order Phase Transitions}},  {\em JCAP} {\bf 1006} (2010)
  028, [\href{http://www.arxiv.org/abs/1004.4187}{{\tt 1004.4187}}].

\bibitem{Megevand2017}
A.~M\'egevand and S.~Ram\'irez, {\it {Bubble nucleation and growth in very
  strong cosmological phase transitions}},  {\em Nucl. Phys.} {\bf B919} (2017)
  74--109, [\href{http://www.arxiv.org/abs/1611.05853}{{\tt 1611.05853}}].

\bibitem{Konstandin:2011dr}
T.~Konstandin and G.~Servant, {\it {Cosmological Consequences of Nearly
  Conformal Dynamics at the TeV scale}},  {\em JCAP} {\bf 1112} (2011) 009,
  [\href{http://www.arxiv.org/abs/1104.4791}{{\tt 1104.4791}}].

\bibitem{Kosowski}
A.~Kosowsky, M.~S. Turner, and R.~Watkins, {\it Gravitational radiation from
  colliding vacuum bubbles},  {\em Phys. Rev. D} {\bf 45} (Jun, 1992)
  4514--4535.

\bibitem{Hindmarsh:2013xza}
M.~Hindmarsh, S.~J. Huber, K.~Rummukainen, and D.~J. Weir, {\it {Gravitational
  waves from the sound of a first order phase transition}},  {\em Phys. Rev.
  Lett.} {\bf 112} (2014) 041301,
  [\href{http://www.arxiv.org/abs/1304.2433}{{\tt 1304.2433}}].

\bibitem{Caprini:2006jb}
C.~Caprini and R.~Durrer, {\it {Gravitational waves from stochastic
  relativistic sources: Primordial turbulence and magnetic fields}},  {\em
  Phys. Rev.} {\bf D74} (2006) 063521,
  [\href{http://www.arxiv.org/abs/astro-ph/0603476}{{\tt astro-ph/0603476}}].

\bibitem{Kuroyanagi:2014qaa}
S.~Kuroyanagi, S.~Tsujikawa, T.~Chiba, and N.~Sugiyama, {\it {Implications of
  the B-mode Polarization Measurement for Direct Detection of Inflationary
  Gravitational Waves}},  {\em Phys. Rev.} {\bf D90} (2014), no.~6 063513,
  [\href{http://www.arxiv.org/abs/1406.1369}{{\tt 1406.1369}}].

\bibitem{Thrane:2013oya}
E.~Thrane and J.~D. Romano, {\it {Sensitivity curves for searches for
  gravitational-wave backgrounds}},  {\em Phys. Rev.} {\bf D88} (2013), no.~12
  124032, [\href{http://www.arxiv.org/abs/1310.5300}{{\tt 1310.5300}}].

\bibitem{Moore:2014lga}
C.~J. Moore, R.~H. Cole, and C.~P.~L. Berry, {\it {Gravitational-wave
  sensitivity curves}},  {\em Class. Quant. Grav.} {\bf 32} (2015), no.~1
  015014, [\href{http://www.arxiv.org/abs/1408.0740}{{\tt 1408.0740}}].

\bibitem{Carrington:1991hz}
M.~E. Carrington, {\it {The Effective potential at finite temperature in the
  Standard Model}},  {\em Phys. Rev.} {\bf D45} (1992) 2933--2944.

\end{thebibliography}\endgroup

\end{document}